\documentclass[11pt]{article}
\usepackage[T1]{fontenc}
\usepackage{amssymb,amsmath,amsfonts}
\usepackage{multirow}
\usepackage{graphicx}
\usepackage[hang,small,sc]{caption}
\usepackage{lscape}
\usepackage{color}
\usepackage[round,sort,comma]{natbib}

\usepackage{hyperref,url}
\usepackage[english]{babel}
\usepackage[autostyle, english = american]{csquotes}
\MakeOuterQuote{"}

\definecolor{darkblue}{rgb}{0,0,0.55}

\numberwithin{equation}{section}
\numberwithin{figure}{section}
\numberwithin{table}{section}

\hypersetup{,colorlinks=true,allcolors=darkblue}

\begin{document}   

\baselineskip 5mm

\thispagestyle{empty}

\begin{center}
{\LARGE Robust Estimation of Loss Models}
\\[2ex]
{\LARGE for Truncated and Censored Severity Data}

\vspace{15mm}

{\large\sc  
Chudamani Poudyal\footnote[1]{
{\sc Corresponding author}: 
Chudamani Poudyal, Ph.D., is
a Visiting Assistant Professor in
the Department of Statistics and Data Science, 
University of Central Florida, 
Orlando, FL 32816, USA. 
~~ {\em e-mail\/}: ~{\tt Chudamani.Poudyal@ucf.edu} 
}}

\vspace{1mm}

{\large\em University of Central Florida}

\vspace{8mm}

{\large\sc
Vytaras Brazauskas\footnote[2]{
Vytaras Brazauskas, Ph.D., ASA,
is a Professor in the Department of Mathematical Sciences,  
University of Wisconsin-Milwaukee, P.O. Box 413, Milwaukee, 
WI 53201, USA. ~~ {\em e-mail\/}: ~{\tt vytaras@uwm.edu}}}  

\vspace{1mm}

{\large\em University of Wisconsin-Milwaukee}


\vspace{20mm}

\copyright \
Copyright of this Manuscript is held by the Authors! 
\end{center}

\vspace{5mm}

\begin{quote}
{\bf\em Abstract\/}.
In this paper, we consider robust estimation of claim severity models 
in insurance, when data are affected by truncation (due to deductibles), 
censoring (due to policy limits), and scaling (due to coinsurance). 
In particular, robust estimators based on the methods of {\em trimmed 
moments\/} ($T$-estimators) and {\em winsorized moments\/} ($W$-estimators) 
are pursued and fully developed. The general definitions of such estimators 
are formulated and their asymptotic properties are investigated. For 
illustrative purposes, specific formulas for $T$- and $W$-estimators of 
the tail parameter of a single-parameter Pareto distribution are derived. 
The practical performance of these estimators is then explored using 
the well-known Norwegian fire claims data. Our results demonstrate that 
$T$- and $W$-estimators offer a robust and computationally efficient 
alternative to the likelihood-based inference for models that are 
affected by deductibles, policy limits, and coinsurance. 

\vspace{4mm}

{\bf\em Keywords\/}. ~Insurance Payments; Loss Models; Robust Estimation;
Trimmed and Winsorized Moments; Truncated and Censored Data.
\end{quote}

\newpage

\baselineskip 7mm
\setcounter{page}{1}

\section{Introduction}
\label{sec:Introduction}

Parametric statistical models for insurance claims severity 
are continuous, right-skewed, and frequently heavy-tailed 
\citep[see][]{MR3890025}. 
The data sets that such models are 
usually fitted to contain outliers that are difficult to identify 
and separate from genuine data. Moreover, due to commonly used 
loss mitigation techniques, the random variables we observe and 
wish to model are affected by data truncation (due to deductibles), 
censoring (due to policy limits), and scaling (due to coinsurance). 
In the current practice, statistical inference for loss models is 
almost exclusively likelihood (MLE) based, which typically results 
in non-robust parameter estimators, pricing models, and risk 
measurements. 

\begin{figure}[hbt!]
\centering
\includegraphics[width=0.98\textwidth] 
{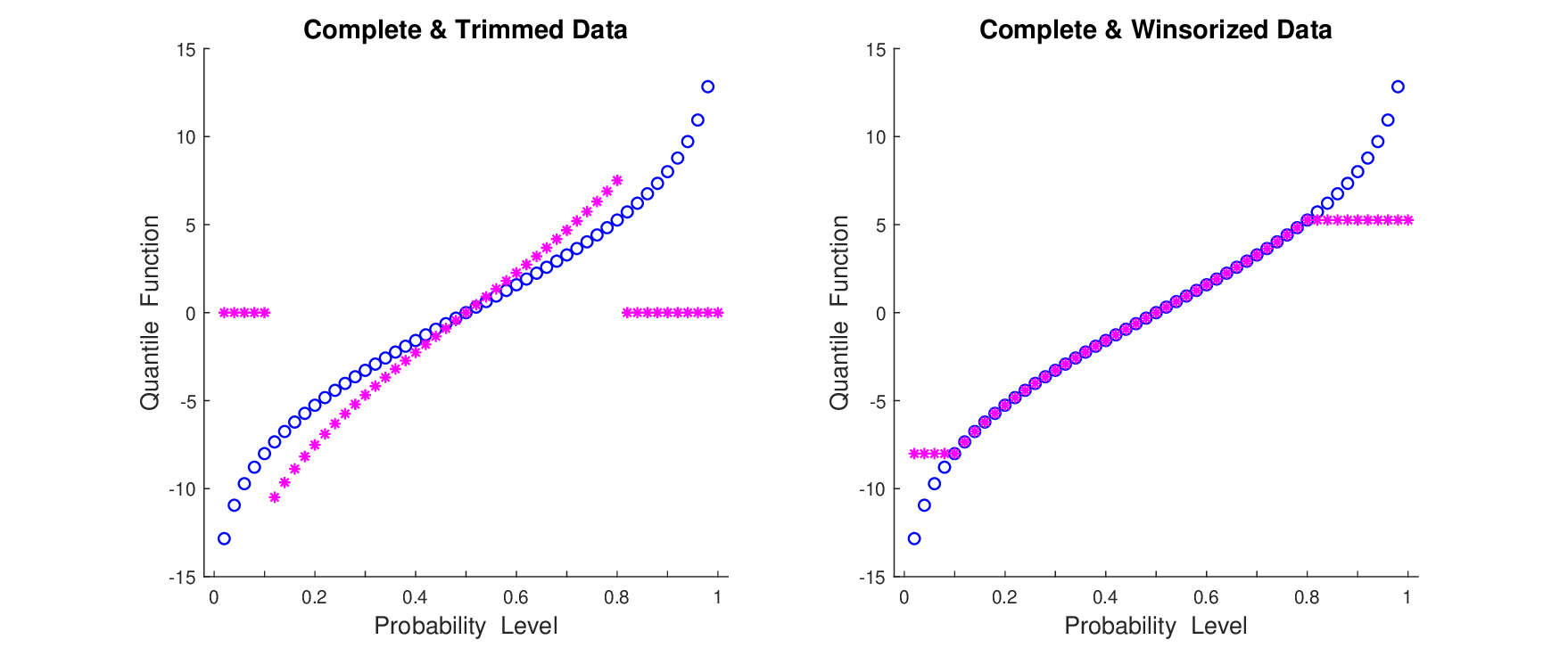}
\caption{Quantile functions of complete data and 
its trimmed and winsorized versions.
Sample size: $n=50$. Trimming/winsorizing proportions: 
10\% (lower) and 20\% (upper).
Complete data marked by `$\circ$' and trimmed/winsorized 
by `$*$'.}
\label{fig:MTM_MWM_Viz}
\end{figure}

Construction of robust actuarial models includes many ideas from 
the mainstream robust statistics literature
\citep[see, e.g.,][]{MR2488795}, but there are 
additional nuances that need to be addressed. 
Namely, actuaries 
have to deal with heavy-tailed and skewed distributions, data 
truncation and censoring, identification and recycling of outliers,
and aggregate loss, just to name a few. 
A number of specialized 
studies addressing some of these issues have been carried out 
in the actuarial literature; see, e.g., 
\cite{k92}, 
\cite{gr93}, 
\cite{MR2035058}, 
\cite{gp00},
\cite{MR1865981}, 
\cite{MR1987777},
and 
\cite{db07}. 
Further, these and other actuarial studies motivated the development 
of two broad classes of robust estimators -- the methods of 
{\em trimmed moments\/} 
\citep[see, e.g.,][]{MR2579571,MR2497558}
and {\em winsorized moments\/} 
\citep[see, e.g.,][]{MR3758788,MR3750626}.
These two approaches, called $T$- and $W$-estimators for short, are sufficiently general and flexible for fitting continuous parametric models based on completely 
observed ground-up loss data. 
In Figure \ref{fig:MTM_MWM_Viz}, we illustrate how $T$ and $W$ methods act on data 
and control the influence of extremes. First of all, notice that 
typical loss mitigation techniques employed in insurance practice 
(e.g., deductibles and policy limits) are closely related to data 
winsorizing or its variants. Secondly, we see that in order to taper 
the effects of rare but high severity claims on parameter estimates, 
data should be ``pre-processed'' using trimming or winsorizing. 
Thenceforth $T$ and $W$ estimates can be found by applying the 
classical {\em method of moments\/}. Note that these initial 
modifications of data have to be taken into account when deriving 
corresponding theoretical moments. This yields an additional benefit. 
Specifically, unlike the parameter estimators based on the standard 
method of moments, which may not exist for heavy-tailed models 
(due to the non-existence of finite moments), theoretical $T$ and 
$W$ moments are always finite. Finally, for trimmed or winsorized 
data, estimation of parameters via the method of moments is not the 
only option. Indeed, one might choose to apply another estimation 
procedure (e.g., properly constructed MLE) and gain similar robustness 
properties. In this paper, however, we focus on rigorous treatment 
of moment-type estimators.

$T$-estimators have been discussed in the operational 
risk literature by 
\cite{oc12}, used in credibility studies by 
\cite{MR3081460}, and further tested in risk measurement 
exercises by 
\cite{MR3904548}. 
Also, the idea of 
trimming has been gaining popularity in modeling extremes 
(see 
\citealp{MR3964266}; 
\citealp{MR4165035}). 
Thus we anticipate the methodology developed in this paper will be 
useful and transferable to all these and other areas of research.

Moreover, besides typical non-robustness of MLE-based inference,
implementation of such procedures on real data is also technically 
challenging 
(see discussions by
\citealp{MR3765327}; 
\citealp{MR3765333}).
This
issue is especially evident when one tries to fit complicated 
multi-parameter models such as finite mixtures of distributions
(see 
\citealp{MR3394072}; 
\citealp{MR3543061}; 
\citealp{MR3724936}).
Thus, 
the primary objective of this paper is to go beyond the complete 
data scenario and develop $T$- and $W$-estimators for insurance 
data affected by the above-mentioned transformations. It will be 
shown that, when properly redesigned, $T$- and $W$-estimators 
can be a robust and computationally efficient alternative to the 
MLE-based inference for claim severity models that are affected 
by deductibles, policy limits, and coinsurance. In particular, 
we will provide the definitions of $T$- and $W$-estimators and 
derive their asymptotic properties such as normality and consistency. 
Specific formulas or estimating equations for a single-parameter 
Pareto (Pareto I) model will be provided. Finally, the practical 
performance of the estimators will be illustrated by fitting 
Pareto I to the well-known Norwegian fire claims data. We will 
use MLE, several $T$- and $W$-estimators, validate the fits, and 
apply the fitted models to price an insurance contract.

The remainder of the paper is organized as follows. In Section 2, 
we describe a series of loss variable (data) transformations, 
starting with complete data, then continuing with truncated and 
censored data, and finishing with two types of insurance payments.
Section 3 uses the data scenarios and models of the previous section
and derives $T$- and $W$-estimators for the parameters of those models. 
Then asymptotic properties of these estimators are established.
In Section 4, we develop specific formulas of the estimators when 
the underlying loss distribution is Pareto I, and compare the asymptotic
relative efficiency of $T$- and $W$-estimators with respect to MLE.
Section 5 is devoted to practical applications of the Pareto I model;
the effects of model fitting on insurance contract pricing are then 
investigated. Finally, concluding remarks are offered in Section 6.

\section{Data and Models}

In this section, we review typical transformations of continuous 
random variables that may be encountered in modeling claim severity. 
For each type of variable transformation, the resulting probability 
density function (pdf), cumulative distribution function (cdf) and 
quantile function (qf) are specified. 

\subsection{Complete Data}

Let us start with the complete data scenario. Suppose the observable 
random variables
\begin{equation}
X_1, \, X_2, \ldots, X_n
\label{data}
\end{equation}
are independent and identically distributed (i.i.d.) and have the 
pdf $f(x)$, cdf $F(x)$, and qf $F^{-1}(v)$. Since loss random 
variables are nonnegative, the support of $f(x)$ is the set 
$\{x: x \geq 0 \}$. 

The complete data scenario is not common when claim severities are 
recorded, but it represents so-called ``ground up'' losses and thus
important to consider. Statistical properties of the ground-up 
variable are of great interest in risk analysis, product design 
(for specifying insurance contract parameters), risk transfer 
considerations, and for other business decisions.

\subsection{Truncated Data}

Data truncation occurs when sample observations are restricted to 
some interval (not necessarily finite), say $(t_1, t_2)$ with 
$t_1 < t_2$. Measurements and even a count of observations outside 
the interval are completely unknown. To formalize this discussion, 
we will say that we observe the i.i.d. data
\begin{equation}
X_1^*, \, X_2^*, \ldots, X_n^*,
\label{tdata}
\end{equation}
where each $X^*$ is equal to the ground-up loss variable $X$, if $X$ 
falls between $t_1$ and $t_2$, and is undefined otherwise. That is, 
$X^*$ satisfies the following conditional event relationship:
\[
X^* \stackrel{d}{=} X \, \big | \, t_1 < X < t_2, 
\]
where $\stackrel{d}{=}$ denotes ``equal in distribution''. Due 
to this relationship, the cdf $F_*$, pdf $f_*$, and qf $F_*^{-1}$ 
of variables $X^*$ are related to $F$, $f$, and $F^{-1}$ (see 
Section 2.1) and given by:
\begin{equation}
F_*( x; \, t_1, t_2 ) = 
\mbox{\bf P} \left[ X \leq x \, \big | \, t_1 < X < t_2 \right] = 
\left\{ 
\begin{array}{cl}
 0, & x \leq t_1; \\[0.5ex] 
\frac{F(x) - F(t_1)}{F(t_2)-F(t_1)}, & t_1 < x < t_2; \\[0.5ex]
 1, & x \geq t_2, \\
\end{array}
\right.
\label{tcdf}
\end{equation}
\begin{equation}
f_*( x; \, t_1, t_2 ) = 
\frac{d}{dx} \Big[ F_*(x; \, t_1, t_2 ) \Big] = 
\left\{
\begin{array}{cl}
\frac{f(x)}{F(t_2)-F(t_1)}, & t_1 < x < t_2; \\[0.5ex]
\mbox{undefined}, & x = t_1, ~ x = t_2; \\[0.25ex]
 0, & \mbox{elsewhere}, \\
\end{array}
\right.
\label{tpdf}
\end{equation}
and
\begin{equation}
F_*^{-1}( v; \, t_1, t_2 ) = 
F^{-1} \big( v F(t_2) + (1-v) F(t_1) \big), 
\quad 
\mbox{for} ~~ 0 \leq v \leq 1.
\label{tqf}
\end{equation}

In industry wide databases such as ORX Loss Data
({\tt managingrisktogether.orx.org}), only losses above some pre-specified 
threshold, say $d$, are collected, which results in {\em left truncated\/} 
data at $d$. Thus, the observations available to the end-user can be viewed 
as a realization of random variables (\ref{tdata}) with $t_1 = d$ and 
$t_2 \rightarrow \infty$. The latter condition slightly simplifies formulas 
(\ref{tcdf})--(\ref{tqf}); one just needs to replace $F(t_2)$ with 1.

\subsection{Censored Data}

There are several versions of data censoring that occur in statistical modeling:
interval censoring (it includes left and right censoring depending on which end
point of the interval is infinite), type I censoring, type II censoring, and
random censoring. For actuarial work, the most relevant type is {\em interval 
censoring\/}. It occurs when complete observations are available within some 
interval, say $(t_1, t_2)$ with $t_1 < t_2$, but data outside the interval is 
only partially known. That is, counts are available but actual values are not. 
To formalize this discussion, we will say that we observe the i.i.d. data
\begin{equation}
X_1^{**}, \, X_2^{**}, \ldots, X_n^{**},
\label{cdata}
\end{equation}
where each $X^{**}$ is equal to the ground-up variable $X$, if $X$ falls between 
$t_1$ and $t_2$, and is equal to the corresponding end-point of the interval if 
$X$ is beyond that point. That is, $X^{**}$ is given by
\[
X^{**} = \min\big\{ \max (t_1, X), \, t_2 \big\} = 
\left\{ 
\begin{array}{cl}
 t_1, & X \leq t_1; \\ 
 X, & t_1 < X < t_2; \\
 t_2, & X \geq t_2. \\
\end{array}
\right.
\]
Due to this relationship, the cdf $F_{**}$, pdf $f_{**}$, and qf 
$F_{**}^{-1}$ of variables $X^{**}$ are related to $F$, $f$, and 
$F^{-1}$ and have the following expressions:
\begin{eqnarray}
F_{**}( x; \, t_1, t_2 ) & = &
\mbox{\bf P} \left[ \min\big\{ \max (t_1, X), \, t_2 \big\} \leq x \right] 
\nonumber
\\
 & = &
\mbox{\bf P} \big[ X \leq x \big] 
\mbox{\large\bf 1} \left\{ t_1 \leq x < t_2 \right\} +
\mbox{\large\bf 1} \left\{ t_2 \leq x \right\} = 
\left\{ 
\begin{array}{cl}
   0, & x < t_1; \\
  F(x), & t_1 \leq x < t_2; \\
   1,  & x \geq t_2, \\
\end{array}
\right.
\label{ccdf}
\end{eqnarray}
where $\mbox{\large\bf 1} \{ \cdot \}$ denotes the indicator function.
Further,
\begin{equation}
F_{**}^{-1}( v; \, t_1, t_2 ) = 
\left\{ 
\begin{array}{cl}
  t_1, & v < F(t_1); \\ 
F^{-1}(v), & F(t_1) \leq v < F(t_2); \\
 t_2, & v \geq F(t_2). \\
\end{array}
\right.
\label{cqf}
\end{equation}
Note that cdf (\ref{ccdf}) is a mixture of continuous cdf $F$ and discrete 
probability mass at $x=t_1$ (with probability $F(t_1)$) and $x=t_2$ (with
probability $1-F(t_2^-)$). This results in a mixed pdf/pmf:
\begin{equation}
f_{**}( x; \, t_1, t_2 ) = 
\left\{
\begin{array}{cl}
 F(t_1), & x = t_1; \\
 f(x), & t_1 < x < t_2; \\
 1 - F(t_2^-), & x = t_2; \\
 0, & \mbox{elsewhere}. \\
\end{array}
\right.
\label{cpdf}
\end{equation}

\subsection{Insurance Payments}

Insurance contracts have coverage modifications that need to be taken into 
account when modeling the underlying loss variable. Usually coverage
modifications such as deductibles, policy limits, and coinsurance are 
introduced as loss control mechanisms so that unfavorable policyholder 
behavioral effects (e.g., adverse selection) can be minimized. There are 
also situations when certain features of the contract emerge naturally 
(e.g., the value of insured property in general insurance is a natural 
upper policy limit). Here we describe two common transformations of 
the loss variable along with the corresponding cdf's, pdf's, and qf's.

Suppose the insurance contract has ordinary deductible $d$, upper policy 
limit $u$, and coinsurance rate $c$ ($0 \leq c \leq 1$). These coverage 
parameters imply that when a loss $X$ is reported, the insurance company 
is responsible for a proportion $c$ of $X$ exceeding $d$, but no more 
than $c(u-d)$. 

Next, if the loss severity $X$ below the deductible $d$ is completely 
unobservable (even its frequency is unknown), then the observed i.i.d.
insurance payments $Y_1, \ldots, Y_n$ can be viewed as realizations of 
{\em left-truncated\/}, {\em right-censored\/}, and 
{\em linearly-transformed\/} $X$ (called {\em payment-per-payment\/} 
variable):
\begin{equation}
Y ~\stackrel{d}{=}~ 
c \left( \min\big\{ X, \, u \big\} - d \right) \, \big | \, X > d ~=~ 
\left\{ 
\begin{array}{cl}
 \mbox{undefined}, & X \leq d; \\
c \left( X-d \right), & d < X < u; \\
c \left( u-d \right), & u \leq X. \\
\end{array}
\right.
\label{p1data}
\end{equation}
We can see that the payment variable $Y$ is a linear transformation of 
a composition of variables $X^*$ and $X^{**}$ (see Sections 2.2 and 2.3). 
Thus, similar to variables $X^*$ and $X^{**}$, its cdf $G_{Y}$, pdf 
$g_{Y}$, and qf $G_{Y}^{-1}$ are also related to $F$, $f$, and $F^{-1}$ 
and given by:
\begin{equation}
G_{Y}( y; \, c, d, u ) = 
\mbox{\bf P} \left[ c \left( 
\min\big\{ X, \, u \big\} - d \right) \leq y \, \big | \, X > d  \right] = 
\left\{ 
\begin{array}{cl}
   0, & y \leq 0; \\[0.5ex] 
 \frac{F(y/c+d) - F(d)}{1-F(d)}, & 0 < y < c(u-d); \\[0.5ex]
   1,  & y \geq c(u-d), \\
\end{array}
\right.
\label{p1cdf}
\end{equation}
\begin{equation}
g_{Y}( y; \, c, d, u ) = 
\left\{
\begin{array}{cl}
 \frac{f(y/c+d)}{c [1-F(d)] }, & 0 < y < c(u-d); \\[1ex]
 \frac{1-F(u^-)}{1-F(d)}, & y = c(u-d); \\[0.75ex]
   0, & \mbox{elsewhere}, \\
\end{array}
\right.
\label{p1pdf}
\end{equation}
and
\begin{equation}
G_{Y}^{-1}( v; \, c, d, u ) = 
\left\{ 
\begin{array}{cl}
 c \left[ F^{-1} \big( v + (1-v) F(d) \big) - d \right], & 
0 \leq v < \frac{F(u)-F(d)}{1-F(d)}; \\[0.75ex]
 c(u-d), & \frac{F(u)-F(d)}{1-F(d)} \leq v \leq 1. \\
\end{array}
\right.
\label{p1qf}
\end{equation}

The scenario that no information is available about $X$ below $d$ is 
likely to occur when modeling is done based on the data acquired from 
a third party (e.g., data vendor). For payment data collected in house,
the information about the number of policies that did not report claims
(equivalently, resulted in a payment of 0) would be available. This minor 
modification yields different payment variables, say $Z_1, \ldots, Z_n$, 
which can be treated as i.i.d. realizations of {\em interval-censored\/} 
and {\em linearly-transformed\/} $X$ (called {\em payment-per-loss\/} 
variable):
\begin{equation}
Z = c \left( \min \big\{ X, u \big\} - \min \big\{ X, d \big\} \right) = 
\left\{ 
\begin{array}{cl}
 0, & X \leq d; \\
c \left( X-d \right), & d < X < u; \\
c \left( u-d \right), & u \leq X. \\
\end{array}
\right.
\label{p2data}
\end{equation}
Again, its cdf $G_{Z}$, pdf $g_{Z}$, and qf $G_{Z}^{-1}$ are related 
to $F$, $f$, and $F^{-1}$ and given by:
\begin{equation}
G_{Z}( z; \, c, d, u ) = 
\mbox{\bf P} \left[ c \left( \min \big\{ X, u \big\} - 
\min \big\{ X, d \big\} \right) \leq z \right] = 
\left\{ 
\begin{array}{cl}
   0, & z < 0; \\[0.25ex]
F(z/c+d), & 0 \leq z < c(u-d); \\[0.25ex]
   1,  & z \geq c(u-d), \\
\end{array}
\right.
\label{p2cdf}
\end{equation}
\begin{equation}
g_{Z}( z; \, c, d, u ) =  
\left\{
\begin{array}{cl}
   F(d), & z = 0; \\[0.25ex]
 f(z/c+d)/c, & 0 < z < c(u-d); \\[0.25ex]
 1 - F(u^-), & z = c(u-d); \\[0.25ex]
    0, & \mbox{elsewhere}, \\
\end{array}
\right.
\label{p2pdf}
\end{equation}
and
\begin{equation}
G_{Z}^{-1}( v; \, c, d, u ) = 
\left\{ 
\begin{array}{cl}
 0, & 0 \leq v \leq F(d); \\[0.25ex]
c \left( F^{-1} (v) - d \right), & F(d) < v < F(u); \\[0.25ex]
 c(u-d), & F(u) \leq v \leq 1. \\
\end{array}
\right.
\label{p2qf}
\end{equation}

\section{{\em T}- and {\em W}-Estimation}

In this section, we first provide definitions of parameter estimators 
obtained by using the Method of Trimmed Moments ($T$-estimators; Section 3.1) 
and the Method of Winsorized Moments ($W$-estimators; Section 3.2) under 
the data scenarios of Sections 2.1--2.4. Then, in Section 3.3, we specify 
the asymptotic distribution of the resulting estimators.
Also, throughout the section we assume that the ground-up losses follow a 
continuous parametric distribution with pdf $f(x \, | \, \boldsymbol{\theta})$ 
and cdf $F(x \, | \, \boldsymbol{\theta})$ which are indexed by $k \ge 1$ 
unknown parameters $\boldsymbol{\theta} = (\theta_1, \ldots, \theta_k)$. 
The goal is to estimate those parameters using $T$- and $W$-estimators 
by taking into account the probabilistic relationships between the cdf 
$F(x \, | \, \boldsymbol{\theta})$ and the distribution function of 
{\em observed\/} data.

\subsection{{\em T}-Estimators}

$T$-estimators are derived by following the standard method-of-moments 
approach, but instead of standard moments we match sample and population 
{\em trimmed\/} ($T$) moments (or their variants). The advantage of such 
an approach over the standard one is that the population $T$ moments always 
exist irrespective of the tail-heaviness of the underlying distribution. 
The following definition lists the formulas of sample and population $T$ 
moments for the data scenarios of Sections 2.1--2.4.

\bigskip

\noindent
{\bf Definition 3.1.} ~{\em For data scenarios and models of Sections 
2.1--2.4, let us denote the sample and population $T$ moments as 
$\widehat{T}_{j}$ and $T_j(\boldsymbol{\theta})$, respectively. If 
$w_{1:n} \leq \cdots \leq w_{n:n}$ is an ordered realization of variables 
(\ref{data}), (\ref{tdata}), (\ref{cdata}), (\ref{p1data}), or (\ref{p2data}) 
with qf denoted $F^{-1}_V(v \, | \, \boldsymbol{\theta})$ (which depending 
upon the data scenario equals to qf $F^{-1}$, (\ref{tqf}), (\ref{cqf}), 
(\ref{p1qf}), or (\ref{p2qf})), then the sample and population $T$ moments,
with the trimming proportions $a$ (lower) and $b$ (upper), have the following 
expressions:
\begin{eqnarray}
\widehat{T}_{j} & = & \frac{1}{n-m_n-m_n^*}
\sum_{i = m_n + 1}^{n - m_n^*}
\big[ h(w_{i:n}) \big]^j, 
\qquad j = 1, \ldots, k,
\label{Ts}
\end{eqnarray}
\begin{eqnarray}
T_j(\boldsymbol{\theta}) & = & \frac{1}{1-a-b} \int_{a}^{1-b} 
\big[ h (F_V^{-1}(v \, | \, \boldsymbol{\theta})) \big]^j \, dv,
\qquad j = 1, \ldots, k.
\label{Tp}
\end{eqnarray}
Under all the data scenarios, the trimming proportions $a$ and $b$ 
and function $h$ are chosen by the researcher. Also, integers $m_n$ and 
$m_{n}^* ~ (0 \le m_n < n - m_n^* \le n)$ are such that $m_n/n \rightarrow a$ 
and $m_n^*/n \rightarrow b$ when $n \rightarrow \infty$. In finite samples, 
the integers $m_n$ and $m_{n}^*$ are computed as $m_n = [n a]$ and 
$m_{n}^* = [n b]$, where $[\cdot]$ denotes the greatest integer part. 
}

\bigskip

\noindent
{\bf Note 3.1.} ~In the original formulation of MTM estimators for complete
data \citep{MR2497558}, the trimming proportions $a$ and 
$b$ and function $h$ were allowed to vary for different $j$, which makes
the technique more flexible. On the other hand, for implementation of MTM 
estimators in practice, such flexibility requires more decisions to be made 
regarding the $a$ and $b$ interaction with each other and for different $h$. 
The follow-up research that used MTMs usually had not varied these constants 
and functions, which seems like a reasonable choice. Therefore, in this paper 
we choose to work with non-varying $a$, $b$, and $h$ for all $j$. 
\hfill $\Box$

\bigskip

\noindent
{\bf Note 3.2.} ~For incomplete data scenarios, possible permutations between 
$a$, $b$ and $F(t_1)$, $F(t_2)$ have to be taken into account. For truncated
data, there is only one possibility: 
$0 \leq F(t_1) \leq a < 1-b \leq F(t_2) \leq 1$.
For censored data, however, it is possible to use part or all of censored 
data in estimation. Thus, we can have six arrangements:
\begin{enumerate}
  \item $0 \leq a < 1-b \leq F(t_1) < F(t_2) \leq 1$. \qquad \qquad 
4. $0 \leq F(t_1) < F(t_2) \leq a < 1-b \leq 1$.
  \item $0 \leq a \leq F(t_1) < 1-b \leq F(t_2) \leq 1$. \qquad \qquad 
5. $0 \leq F(t_1) \leq a < F(t_2) \leq 1-b \leq 1$.
  \item $0 \leq a \leq F(t_1) < F(t_2) \leq 1-b \leq 1$. \qquad \qquad 
6. $0 \leq F(t_1) \leq a < 1-b \leq F(t_2) \leq 1$.
\end{enumerate}
Among these, the case 6 ($0 \leq F(t_1) \leq a < 1-b \leq F(t_2) \leq 1$) 
makes most sense because it uses the available data in the most effective way. 
For the sake of completeness, however, we will investigate the other cases as 
well (see Section 4). Note that the insurance payments $Y$ and $Z$ are special 
(mixed) cases of truncated and censored data and thus will possess similar 
properties. Moreover, the $T$-estimators based on case 6 will be resistant 
to outliers, i.e., observations that are inconsistent with the assumed model 
and most likely appearing at the boundaries $t_1$ and $t_2$.
\hfill $\Box$

\bigskip

\noindent
{\bf Note 3.3.} ~In view of Notes 3.1 and 3.2, the $T$-estimators with $a>0$
and $b>0$ ($0 \leq F(t_1) \leq a < 1-b \leq F(t_2) \leq 1$) are globally 
robust with the {\em lower\/} and {\em upper\/} breakdown points given by
$\mbox{\sc lbp} = a$ and $\mbox{\sc ubp} = b$, respectively. The robustness
of such estimators against small or large outliers comes from the fact that
in the computation of estimates the influence of the order statistics with
the index less than $n \times \mbox{\sc lbp}$ or higher than 
$n \times (1- \mbox{\sc ubp})$ is limited. For more details on {\sc lbp} 
and {\sc ubp}, see 
\cite{MR1989836} 
and 
\cite{MR1987777}.
\hfill $\Box$

\medskip

\subsection{{\em W}-Estimators}

$W$-estimators are derived by following the standard method-of-moments 
approach, but instead of standard moments we match sample and population 
{\em winsorized\/} ($W$) moments (or their variants). Similar to $T$-estimators, 
the population $W$ moments also always exist. The following definition lists 
the formulas of sample and population $W$ moments for the data scenarios of 
Sections 2.1--2.4.

\bigskip

\noindent
{\bf Definition 3.2.} ~{\em For data scenarios and models of Sections 
2.1--2.4, let us denote the sample and population $W$ moments as 
$\widehat{W}_{j}$ and $W_j(\boldsymbol{\theta})$, respectively. If 
$w_{1:n} \leq \cdots \leq w_{n:n}$ is an ordered realization of variables 
(\ref{data}), (\ref{tdata}), (\ref{cdata}), (\ref{p1data}), or (\ref{p2data}) 
with qf denoted $F^{-1}_V(v \, | \, \boldsymbol{\theta})$ (which depending 
upon the data scenario equals to qf $F^{-1}$, (\ref{tqf}), (\ref{cqf}), 
(\ref{p1qf}), or (\ref{p2qf})), then the sample and population $W$ moments,
with the winsorizing proportions $a$ (lower) and $b$ (upper), have the 
following expressions:
\begin{eqnarray}
\widehat{W}_{j} & = & \frac{1}{n} 
\left[ 
m_n \big[ h(w_{m_n+1:n}) \big]^j +
\sum_{i=m_n+1}^{n-m_n^*} \big[ h(w_{i:n}) \big]^j +
m_n^* \big[ h(w_{n - m_n^*:n}) \big]^j 
\right], 
\label{Ws}
\\[1ex]
W_j(\boldsymbol{\theta}) & = & 
a \big[ h(F_V^{-1}(a \, | \, \boldsymbol{\theta})) \big]^j +
\int_{a}^{1-b} \big[ h(F_V^{-1}(v \, | \, \boldsymbol{\theta})) \big]^j \, dv +
b \big[ h(F_V^{-1}(1-b \, | \, \boldsymbol{\theta})) \big]^j,
\label{Wp}
\end{eqnarray}
where $j = 1, \ldots, k$, the winsorizing proportions $a$, $b$ and 
function $h$ are chosen by the researcher, and integers $m_n$, $m_{n}^* $ 
are defined and computed the same way as in Definition 3.1.
}

\bigskip

\noindent
{\bf Note 3.4.} ~In the original formulation of MWM estimators for complete 
data \citep{MR3758788}, the winsorizing proportions $a$ and 
$b$ and function $h$ were allowed to vary for different $j$. Based on the 
arguments similar to those made in Note 3.1, in this paper we will choose 
the same $a$, $b$, and $h$ for all $j$. Further, the focus will be on the
case when $a$ and $1-b$ fall within the interval $[F(t_1); \, F(t_2)]$:
$0 \leq F(t_1) \leq a < 1-b \leq F(t_2) \leq 1$. Finally, the breakdown 
points of $W$-estimators are identical to those of $T$-estimators, i.e., 
$\mbox{\sc lbp} = a$ and $\mbox{\sc ubp} = b$.
\hfill $\Box$

\medskip

\subsection{Asymptotic Properties}

In this section, the asymptotically normal distributions for the $T$- and 
$W$-estimators of Sections 3.1--3.2 are specified. It follows immediately
from the parametric structure of those asymptotic distributions that all 
the estimators under consideration are {\em consistent\/}. Throughout 
the section the notation $\cal{AN}$ is used to denote ``asymptotically 
normal''.

\subsubsection{{\em T}-Estimators}

$T$-estimators are found by matching sample $T$-moments (\ref{Ts}) with 
population $T$-moments (\ref{Tp}) for $j = 1, \ldots, k$, and then solving 
the system of equations with respect to $\theta_1, \ldots, \theta_k$.
The obtained solutions, which we denote by 
$\widehat{\theta}_j = s_j(\widehat{T}_{1}, \ldots, \widehat{T}_{k})$,
$1 \leq j \leq k$, are, by definition, the $T$-estimators of 
$\theta_1, \ldots, \theta_k$. Note that the functions $s_j$ are such that
$\theta_j = s_j(T_1(\boldsymbol{\theta}), \ldots, T_k(\boldsymbol{\theta}))$.

The asymptotic distribution of these estimators for complete data has been 
derived by \cite{MR2497558}. It also follows from a more 
general theorem established by 
\citet[Note 2.4]{MR3758788},
which relies on central limit theory of $L$-statistics 
\citep{MR0203874}. 
The following theorem generalizes those results to all data scenarios 
of Sections 2.1--2.4.

\bigskip

\noindent
{\bf Theorem 3.1.} ~{\em Suppose an i.i.d. realization of variables 
(\ref{data}), (\ref{tdata}), (\ref{cdata}), (\ref{p1data}), or (\ref{p2data}) 
has been generated by cdf $F_V(v \, | \, \boldsymbol{\theta})$ which 
depending upon the data scenario equals to cdf $F$, (\ref{tcdf}), 
(\ref{ccdf}), (\ref{p1cdf}), or (\ref{p2cdf}), respectively. Let 
$\widehat{\boldsymbol{\theta}}_{\mbox{\tiny T}} = 
\left( \widehat{\theta}_1, \ldots, \widehat{\theta}_k \right) = 
\left( s_1(\widehat{T}_{1}, \ldots, \widehat{T}_{k}), \ldots,
s_k(\widehat{T}_{1}, \ldots,\widehat{T}_{k}) \right)$ denote 
a $T$-estimator of $\boldsymbol{\theta}$. Then
\[
\widehat{\boldsymbol{\theta}}_{\mbox{\tiny T}} = 
\left( \widehat{\theta}_1, \ldots, \widehat{\theta}_k \right) ~~is~~ 
{\cal{AN}}
\left( 
\big( \theta_1, \ldots, \theta_k \big), \, \frac{1}{n} \, 
\mathbf{D}_t \boldsymbol{\Sigma}_t \mathbf{D}_t'
\right),
\]
where $\mathbf{D}_t := \big[ d_{ij} \big]_{i,j=1}^{k}$ is the Jacobian 
of the transformations $s_1, \ldots, s_k$ evaluated at 
$\big( T_1(\boldsymbol{\theta}), \ldots, T_k(\boldsymbol{\theta}) \big)$
and $\mathbf{\Sigma}_t := \big[ \sigma^2_{ij} \big]_{i,j=1}^{k}$ is 
the covariance-variance matrix with the entries
\[
\sigma^2_{ij} = \frac{1}{(1-a-b)(1-a-b)}
\int_{a}^{1-b} \int_{a}^{1-b}
\big( \min \{ v, w \} - v w \big) \;
\mbox{d} \left[ h \big( F_V^{-1}(v) \big) \right]^j \,
\mbox{d} \left[ h \big( F_V^{-1}(w) \big) \right]^i.
\]
}

\noindent
{\bf Proof.} ~For complete data, generated by (\ref{data}) and with 
the assumption that $F_{V} \equiv F$ is continuous, see 
\cite{MR2497558} or \citet[Note 2.4]{MR3758788}.

For truncated data, generated by (\ref{tdata}), the cdf $F_{*}$ given 
by \eqref{tcdf} is still continuous and hence the results established 
for complete data can be directly applied to $F_{*}$.

For the remaining data scenarios, generated by (\ref{cdata}), (\ref{p1data}), 
or (\ref{p2data}), the qf $F_{V}^{-1}$ is not smooth and the functions 
$
H_{j}
=
\left[ 
h \circ F_{V}^{-1}
\right]^{j}, \ 
j = 1,2, \ldots, k
$
have points of non-differentiability 
(see Lemma A.1 in \citealp{MR3758788}).
The set of such points, however, has probability zero, which means that 
the cdf's $F_{**}$, $G_{Y}$, and $G_{Z}$ are {\em almost everywhere\/} 
continuous under the Borel probability measures induced by $F_{**}$, 
$G_{Y}$, and $G_{Z}$ 
\citep[see, e.g.,][Theorem 1.16]{MR1681462}. 
Therefore, $H'_j$ shall be replaced with $0$ whenever
it is not defined; 
see \citet[Assumption A$^*$]{MR0203874}.
\hfill $\Box$

\medskip

\noindent
{\bf Note 3.5.} ~Theorem 3.1 states that $T$-estimators for the parameters 
of loss models considered in this paper are asymptotically unbiased with 
the entries of the covariance-variance matrix diminishing at the rate $1/n$. 
Using these properties in conjunction with the multidimensional Chebyshev's 
inequality it is a straightforward exercise to establish the fact that 
$T$-estimators are consistent.
\hfill $\Box$

\medskip

\subsubsection{{\em W}-Estimators}

$W$-estimators are found by matching sample $W$-moments (\ref{Ws}) with 
population $W$-moments (\ref{Wp}) for $j = 1, \ldots, k$, and then solving 
the system of equations with respect to $\theta_1, \ldots, \theta_k$.
The obtained solutions, which we denote by 
$\widehat{\theta}_j = r_j(\widehat{W}_{1}, \ldots, \widehat{W}_{k})$,
$1 \leq j \leq k$, are, by definition, the $W$-estimators of 
$\theta_1, \ldots, \theta_k$. Note that the functions $r_j$ are such that
$\theta_j = r_j(W_1(\boldsymbol{\theta}), \ldots, W_k(\boldsymbol{\theta}))$.

The asymptotic distribution of these estimators for complete data has been 
established by 
\citet[Theorem 2.1 and Lemma A.1.]{MR3758788}.
The following theorem summarizes the asymptotic distribution of 
$W$-estimators to all data scenarios of Section 2.

\bigskip

\noindent
{\bf Theorem 3.2.} ~{\em Suppose an i.i.d. realization of variables 
(\ref{data}), (\ref{tdata}), (\ref{cdata}), (\ref{p1data}), or (\ref{p2data}) 
has been generated by cdf $F_V(v \, | \, \boldsymbol{\theta})$ which 
depending upon the data scenario equals to cdf $F$, (\ref{tcdf}), 
(\ref{ccdf}), (\ref{p1cdf}), or (\ref{p2cdf}), respectively. Let 
$\widehat{\boldsymbol{\theta}}_{\mbox{\tiny W}} = 
\left( \widehat{\theta}_1, \ldots, \widehat{\theta}_k \right) = 
\left( r_1(\widehat{W}_{1}, \ldots, \widehat{W}_{k}), \ldots,
r_k(\widehat{W}_{1}, \ldots, \widehat{W}_{k}) \right)$ denote 
a $W$-estimator of $\boldsymbol{\theta}$. Then
\[
\widehat{\boldsymbol{\theta}}_{\mbox{\tiny W}} = 
\left( \widehat{\theta}_1, \ldots, \widehat{\theta}_k \right) ~~is~~ 
{\cal{AN}}
\left( 
\big( \theta_1, \ldots, \theta_k \big), \, \frac{1}{n} \, 
\mathbf{D}_w \boldsymbol{\Sigma}_w \mathbf{D}_w'
\right),
\]
where $\mathbf{D}_w := \big[ d_{ij} \big]_{i,j=1}^{k}$ is the Jacobian 
of the transformations $r_1, \ldots, r_k$ evaluated at 
$\big( W_1(\boldsymbol{\theta}), \ldots, W_k(\boldsymbol{\theta}) \big)$
and $\mathbf{\Sigma}_w := \big[ \sigma^2_{ij} \big]_{i,j=1}^{k}$ is 
the covariance-variance matrix with the entries 
\[
\sigma^2_{ij} = \widehat{A}_{i,j}^{(1)} + \widehat{A}_{i,j}^{(2)} + 
\widehat{A}_{i,j}^{(3)} + \widehat{A}_{i,j}^{(4)},
\]
where the terms $\widehat{A}_{i,j}^{(m)}, \; m = 1, \ldots, 4$, are 
specified in \citet[Lemma A.1]{MR3758788}.
}

\smallskip

\noindent
{\bf Proof.} ~The proof can be established by following the same arguments 
as in Theorem 3.1.
\hfill $\Box$

\medskip

\noindent
{\bf Note 3.6.} ~Similar to the discussion of Note 3.5, the asymptotic
normality statement of this theorem implies that $W$-estimators are 
consistent.
\hfill $\Box$

\section{Analytic Examples: Pareto I}

In this section, we first derive MLE, $T$- and $W$-estimators for 
the tail parameter of a single-parameter Pareto distribution, 
abbreviated as Pareto I, when the observed data is either in the 
form of insurance payments $Y$, defined by (\ref{p1data}), or $Z$, 
defined by (\ref{p2data}). 
The corresponding MLE and $T$-estimators for
lognormal distribution has recently been 
investigated by \cite{MR4263275}.
Note that Pareto I is the distribution 
of the ground-up variable $X$. The cdf, pdf, and qf of Pareto I 
are defined as follows:
\begin{eqnarray}
\mbox{\sc cdf:} \qquad 
F(x) & = & 1 - (x_0/x)^{\alpha}, \qquad x > x_0,
\label{PaIcdf}
\\
\mbox{\sc pdf:} \qquad 
f(x) & = & (\alpha/x_0) (x_0/x)^{\alpha + 1}, \qquad x > x_0,
\label{PaIpdf}
\\
\mbox{\sc qf:} \quad ~
F^{-1}(v) & = & x_0 (1-v)^{-1/\alpha}, \qquad 0 \leq v \leq 1,
\label{PaIqf}
\end{eqnarray}
where $\alpha > 0$ is the shape (tail) parameter and $x_0 > 0$ is 
a known constant.

Then, the definitions of the estimators are complemented with their
asymptotic distributions. Using the asymptotic normality results, 
we evaluate the asymptotic relative efficiency (ARE) of the $T$- 
and $W$-estimators with respect to the MLE:
\[
\mbox{ARE$\big($Q, MLE$\big)$} ~=~ 
\frac{\mbox{asymptotic variance of MLE estimator}}
{\mbox{asymptotic variance of Q estimator}} \, ,
\]
where Q represents $T$ or $W$ estimator. Since for Pareto I the 
asymptotic variance of MLE reaches the Cram{\'{e}}r-Rao lower bound, 
the other estimators' efficiency will be between 0 and 1. Estimators 
with AREs close to 1 are preferred.

Also, for the complete data scenario, formulas of 
$\widehat{\alpha}_{\mbox{\tiny MLE}}$ and 
$\widehat{\alpha}_{\mbox{\em\tiny T}}$
are available in \cite{MR2497558}. Derivations for 
the other data scenarios of Section 2 (truncated and censored data) 
are analogous to the ones presented in this section and thus will 
be skipped.

\subsection{MLE}

\subsubsection{Payments {\em Y}}

If $y_1, \ldots, y_n$ is a realization of variables (\ref{p1data}) with 
pdf (\ref{p1pdf}) and cdf (\ref{p1cdf}), where $F$ and $f$ are given by 
(\ref{PaIcdf}) and (\ref{PaIpdf}), respectively, then the log-likelihood 
function can be specified by following standard results presented in
\citet[Chapter 11]{MR3890025}:
\begin{eqnarray*}
{\cal{L}}_{P_Y} \big( \alpha \, \big| \, y_1, \ldots, y_n \big) & = & 
 \sum_{i=1}^n \log \big[ f(y_i/c+d) / c \big]
\mbox{\large\bf 1} \{ 0 < y_i < c(u-d) \}
\\
 & & -~ n \log \big[ 1 - F(d) \big] ~+~
\log \big[ 1 - F(u^-) \big] \sum_{i=1}^n \mbox{\large\bf 1} \{ y_i = c(u-d) \}
\\
 & = &
\sum_{i=1}^n \left[ \log \left( \frac{\alpha}{c x_0} \right) - 
(\alpha+1) \log \left( \frac{y_i/c+d}{x_0} \right) \right]
\mbox{\large\bf 1} \{ 0 < y_i < c(u-d) \}
\\
 & & -~ \alpha n \log (x_0/d) ~+~ \alpha \log (x_0/u)
\sum_{i=1}^n \mbox{\large\bf 1} \{ y_i = c(u-d) \},
\end{eqnarray*}
where $\mbox{\large\bf 1} \{ \cdot \}$ denotes the indicator function.
Straightforward maximization of ${\cal{L}}_{P_Y}$ yields an explicit formula 
of the MLE of $\alpha$:
\begin{equation}
\widehat{\alpha}_{\mbox{\tiny MLE}} = 
\frac{\sum_{i=1}^n \mbox{\large\bf 1} \{ 0 < y_i < c(u-d) \}}
{\sum_{i=1}^n \log ( y_i/(cd) + 1 ) \mbox{\large\bf 1} \{ 0 < y_i < c(u-d) \} +
\log (u/d) \sum_{i=1}^n \mbox{\large\bf 1} \{ y_i = c(u-d) \}}
\label{PaImle}
\end{equation}
The asymptotic distribution of $\widehat{\alpha}_{\mbox{\tiny MLE}}$ follows 
from standard results for MLEs 
\citep[see, e.g.,][Section 4.2]{MR595165}. 
In this case, the Fisher information matrix has a single entry: 
\[
I_{11} ~=~ - \mbox{\bf E} 
\left[  
\frac{\partial^2 \log g_Y(Y \, | \, \alpha)}{\partial \alpha^2} 
\right]
 ~=~ - \mbox{\bf E} 
\left[ - \frac{1}{\alpha^2} \mbox{\large\bf 1} \{ 0 < Y < c(u-d) \} \right]
 ~=~
\frac{1}{\alpha^2} \big[ 1 - ( d/u )^{\alpha} \big].
\]
Hence, the estimator $\widehat{\alpha}_{\mbox{\tiny MLE}}$, defined by 
(\ref{PaImle}), has the following asymptotic distribution:
\begin{equation}
\widehat{\alpha}_{\mbox{\tiny MLE}} ~~is~~ 
{\cal{AN}}
\left( 
\alpha, \, \frac{1}{n} \, \frac{\alpha^2}{1 - ( d/u )^{\alpha}}
\right).
\label{PaImle-an}
\end{equation}
A few observations can be made from this result. First, the coinsurance factor 
$c$ has no effect on (\ref{PaImle-an}). Second, the corresponding result for the 
complete data scenario is obtained when there is no deductible (i.e., $d=x_0$) 
and no policy limit (i.e., $u \rightarrow \infty$). Third, if $u \rightarrow \infty$,
then the asymptotic properties of $\widehat{\alpha}_{\mbox{\tiny MLE}}$ remain 
equivalent to those of the complete data case irrespective of the choice of 
$d$ ($d < \infty$). Also, notice that (\ref{PaImle-an}) implies that 
$\widehat{\alpha}_{\mbox{\tiny MLE}}$ is a consistent and efficient estimator.

\subsubsection{Payments {\em Z}}

If $z_1, \ldots, z_n$ is a realization of variables (\ref{p2data}) with 
pdf (\ref{p2pdf}) and cdf (\ref{p2cdf}), where $F$ and $f$ are given by 
(\ref{PaIcdf}) and (\ref{PaIpdf}), respectively, then the log-likelihood 
function can be specified by following standard results presented in
\citet[Chapter 11]{MR3890025}:
\begin{eqnarray}
{\cal{L}}_{P_Z} \big( \alpha \, \big| \, z_1, \ldots, z_n \big) & = & 
\log \big[ F(d) \big] \sum_{i=1}^n \mbox{\large\bf 1} \{ z_i = 0 \}
~+~ \log \big[ 1 - F(u^-) \big] 
\sum_{i=1}^n \mbox{\large\bf 1} \{ z_i =  c(u-d) \}
\nonumber
\\
 & & +
\sum_{i=1}^n \log \big[ f(z_i /c + d) / c \big]
\mbox{\large\bf 1} \{ 0 < z_i < c(u-d) \}
\nonumber
\\
 & = & 
\log \big[ 1 - (x_0/d)^{\alpha} \big] \sum_{i=1}^n \mbox{\large\bf 1} \{ z_i = 0 \}
~+~ \alpha \log (x_0/u) \sum_{i=1}^n \mbox{\large\bf 1} \{ z_i =  c(u-d) \}
\nonumber
\\
 & & +
\sum_{i=1}^n \left[ \log \left( \frac{\alpha}{cx_0} \right) - 
(\alpha+1) \log \left( \frac{z_i/c+d}{x_0} \right) \right]
\mbox{\large\bf 1} \{ 0 < z_i < c(u-d) \}. \qquad ~
\label{PaImle2}
\end{eqnarray}
It is clear from the expression of ${\cal{L}}_{P_Z}$ that it has to be maximized
numerically. Suppose that a unique solution for maximization of (\ref{PaImle2}) 
with respect to $\alpha$ is found, and let us denote it 
$\widehat{\widehat{\alpha}}_{\mbox{\tiny MLE}}$.

Further, the asymptotic distribution of 
$\widehat{\widehat{\alpha}}_{\mbox{\tiny MLE}}$ 
follows from standard results for MLEs 
\citep[see, e.g.,][Section 4.2]{MR595165}. 
In this case, the single entry of the Fisher 
information matrix is
\begin{eqnarray*}
I_{11} & = & - \mbox{\bf E} 
\left[  
\frac{\partial^2 \log g_Z(Z \, | \, \alpha)}{\partial \alpha^2} 
\right]
 ~=~ - \mbox{\bf E} 
\left[ 
- \frac{(x_0/d)^{\alpha} \log^2 (x_0/d)}{(1 - (x_0/d)^{\alpha})^2} 
\mbox{\large\bf 1} \{ Z = 0 \}
- \frac{1}{\alpha^2} \mbox{\large\bf 1} \{ 0 < Z < c(u-d) \} 
\right]
\\[1ex]
 & = & 
\alpha^{-2} \left[ 
\frac{(x_0/d)^{\alpha}}{1-(x_0/d)^{\alpha}}
\log^2 \left[ (x_0/d)^{\alpha} \right] + 
(x_0/d)^{\alpha} - (x_0/u)^{\alpha} \right] .
\end{eqnarray*}
Hence, the estimator $\widehat{\widehat{\alpha}}_{\mbox{\tiny MLE}}$, 
found by numerically maximizing (\ref{PaImle2}), has the following 
asymptotic distribution:
\begin{equation}
\widehat{\widehat{\alpha}}_{\mbox{\tiny MLE}} ~~is~~ 
{\cal{AN}}
\left( 
\alpha, \, \frac{\alpha^2}{n} \, 
\left[ 
\frac{(x_0/d)^{\alpha}}{1-(x_0/d)^{\alpha}}
\log^2 \left[ (x_0/d)^{\alpha} \right] + 
(x_0/d)^{\alpha} - (x_0/u)^{\alpha} \right]^{-1}
\right).
\label{PaImle2-an}
\end{equation}
Here, we will again emphasize several points. First, as in Section 4.1.1, 
the coinsurance factor $c$ has no effect on (\ref{PaImle2-an}). Second, 
the corresponding result for the complete data scenario is obtained 
when there is no deductible (to eliminate $d$ from (\ref{PaImle2-an}), 
take the limit as $d \rightarrow x_0$) and no policy limit (i.e., 
$u \rightarrow \infty$). Third, (\ref{PaImle2-an}) implies that 
$\widehat{\widehat{\alpha}}_{\mbox{\tiny MLE}}$ is a consistent 
and efficient estimator.

\subsection{{\em T}-Estimators}

\subsubsection{Payments {\em Y}}

Let $y_{1:n} \leq \cdots \leq y_{n:n}$ denote an ordered realization of 
variables (\ref{p1data}) with qf (\ref{p1qf}), where $F$ and $F^{-1}$ 
are given by (\ref{PaIcdf}) and (\ref{PaIqf}), respectively. Since 
Pareto I has only one unknown parameter, we need only one $T$ moment 
equation to estimate it. Also, since payments $Y$ are left-truncated 
and right-censored, it follows from Note 3.2 that only the last three 
permutations between the trimming proportions $a$, $b$ and $F(t_1)$, 
$F(t_2)$ are possible (i.e., $a$ cannot be below $F(t_1)$). That is, 
after converting $t_1$ and $t_2$ into the notation involving $c$, $d$, 
and $u$, we get from (\ref{p1qf}) the following arrangements:

\medskip

Case 1: ~$0 < \frac{F(u)-F(d)}{1-F(d)} \leq a < 1-b \leq 1$ 
~(estimation based on censored data only).

Case 2: ~$0 \leq a < \frac{F(u)-F(d)}{1-F(d)} \leq 1-b \leq 1$ 
~(estimation based on observed and censored data).

Case 3: ~$0 \leq a < 1-b \leq \frac{F(u)-F(d)}{1-F(d)} \leq 1$ 
~(estimation based on observed data only).

\medskip

\noindent
In all these cases, the sample $T$ moments (\ref{Ts}) can be easily
computed by first estimating the probability $[F(u)-F(d)]/[1-F(d)]$ 
as $n^{-1} \sum_{i=1}^n \mbox{\large\bf 1} \{ 0 < y_i < c(u-d) \}$,
then selecting $a$, $b$, and finally choosing $h_Y(y) = \log(y/(cd)+1)$. 
Note that $c$ and $d$ are known constants, and the logarithmic 
transformation will linearize the qf $F^{-1}$ in terms of $\alpha^{-1}$
(at least for the observed data part). With these choices in mind, 
let us examine what happens to the population $T$ moments (\ref{Tp}) 
under the cases 1--3. The following steps can be easily verified:
\begin{eqnarray*}
(1-a-b) \, T_{1(y)} (\alpha) & = & \int_a^{1-b} 
h_Y \left( G_Y^{-1} (v \, | \, \alpha) \right) \, dv
~=~ \int_a^{1-b} 
\log \left( \frac{G_Y^{-1} (v \, | \, \alpha)}{cd} + 1 \right) \, dv
\\[2ex]
 & = & \int_a^{1-b} 
\Bigg[
\log \left( \frac{1}{d} \, F^{-1} \Big( v + (1-v) F(d) \Big) \right)
\mbox{\large\bf 1} \left\{ 0 \leq v < \frac{F(u)-F(d)}{1-F(d)} \right\}
\\[1ex]
 & & ~+~ 
\log \left( u/d \right)
\mbox{\large\bf 1} \left\{ \frac{F(u)-F(d)}{1-F(d)} \leq v \leq 1 \right\}
\Bigg]
\, dv
\\[2ex]
 & = & 
\left\{
 \begin{array}{cl}
(1-a-b) \log (u/d), & \mbox{~ Case 1}; \\[1ex]
\alpha^{-1}
\Big[
(1-a) \big( 1 - \log (1-a) \big) +
b \log \left( d/u \right)^{\alpha} - \left( d/u \right)^{\alpha}
\Big], &
\mbox{~ Case 2}; \\[2ex]
\alpha^{-1} 
\Big[ 
(1-a) \big( 1 - \log (1-a) \big) - b \big( 1 - \log b \big)
\Big], &
\mbox{~ Case 3}. \\
 \end{array}
\right.
\end{eqnarray*}
It is clear from these expressions that estimation of $\alpha$ is impossible 
in Case 1 because there is no $\alpha$ in the formula of $T_{1(y)}(\alpha)$. 
In Case 2, $\alpha$ has to be estimated numerically by solving the following 
equation:
\begin{equation}
\alpha^{-1} \Big[ (1-a) \big( 1 - \log (1-a) \big) +
b \log \left( d/u \right)^{\alpha} - \left( d/u \right)^{\alpha} \Big]
~=~
(1-a-b) \widehat{T}_{1(y)},
\label{PaIyT2}
\end{equation}
where $\widehat{T}_{1(y)} = (n-m_n-m_n^*)^{-1}
\sum_{i = m_n + 1}^{n - m_n^*} \log( y_{i:n}/(cd) + 1 )$.
Suppose a unique solution of (\ref{PaIyT2}) with respect to $\alpha$ is found.
Let us denote it $\widehat{\alpha}_{\mbox{\tiny $T$}}^{(2)}$ and remember that
it is a function of $\widehat{T}_{1(y)}$, say $s_1^{(2)}(\widehat{T}_{1(y)})$. 
Finally, if Case 3 is chosen, then we have an explicit formula for a 
$T$-estimator of 
$\alpha$:
\begin{equation}
\widehat{\alpha}_{\mbox{\tiny $T$}}^{(3)} = 
\frac{I_t(a,1-b)}{(1-a-b) \widehat{T}_{1(y)}} ~=:~ s_1^{(3)}(\widehat{T}_{1(y)}),
\label{PaIyT3}
\end{equation}
where $I_t(a,1-b) := - \int_a^{1-b} \log (1-v) \, dv = (1-a) (1 - \log (1-a)) - 
b (1 - \log b)$ and the sample $T$ moment $\widehat{T}_{1(y)}$ is computed 
as before; see (\ref{PaIyT2}).

Next, we will specify the asymptotic distributions and compute AREs of 
$\widehat{\alpha}_{\mbox{\tiny $T$}}^{(2)} = s_1^{(2)}(\widehat{T}_{1(y)})$ and 
$\widehat{\alpha}_{\mbox{\tiny $T$}}^{(3)} = s_1^{(3)}(\widehat{T}_{1(y)})$.
The asymptotic distributions of $\widehat{\alpha}_{\mbox{\tiny $T$}}^{(2)}$
and $\widehat{\alpha}_{\mbox{\tiny $T$}}^{(3)}$ follow from Theorem 3.1. 
In both cases, the Jacobian $\mathbf{D}_t$ and the covariance-variance 
matrix $\mathbf{\Sigma}_t$ are scalar. Denoting $d_{11}^{(2)}$ and
$d_{11}^{(3)}$ the Jacobian entries for Cases 2 and 3, respectively, 
we get the following expressions:
\begin{eqnarray*}
d_{11}^{(2)} & = & \frac{\partial \widehat{\alpha}_{\mbox{\tiny $T$}}^{(2)}}
{\partial \widehat{T}_{1(y)}} \Bigg|_{\widehat{T}_{1(y)} = T_{1(y)}} ~=~
\frac{\partial s_1^{(2)} (\widehat{T}_{1(y)})}
{\partial \widehat{T}_{1(y)}} \Bigg|_{\widehat{T}_{1(y)} = T_{1(y)}} 
\\[1ex]
 & = & \frac{(1-a-b) \alpha^2}{(d/u)^{\alpha} \left( 1 - \log (d/u)^{\alpha} \right)
- (1-a) \left( 1 - \log (1-a) \right)} ~=~ 
- \frac{(1-a-b) \alpha^2}{I_t(a,1-(d/u)^{\alpha})},
\\[2ex]
d_{11}^{(3)} & = & \frac{\partial \widehat{\alpha}_{\mbox{\tiny $T$}}^{(3)}}
{\partial \widehat{T}_{1(y)}} \Bigg|_{\widehat{T}_{1(y)} = T_{1(y)}} ~=~
\frac{\partial s_1^{(3)} (\widehat{T}_{1(y)})}
{\partial \widehat{T}_{1(y)}} \Bigg|_{\widehat{T}_{1(y)} = T_{1(y)}} ~=~ 
- \frac{(1-a-b) \alpha^2}{I_t(a,1-b)}.
\end{eqnarray*}
Note that $d_{11}^{(2)}$ is found by implicitly differentiating (\ref{PaIyT2}).
Further, denoting $\sigma_{11(2)}^{2}$ and $\sigma_{11(3)}^{2}$ the  
$\mathbf{\Sigma}_t$ entries for Cases 2 and 3, respectively, we get 
the following expressions:
\begin{eqnarray*}
(1-a-b)^2 \sigma_{11(2)}^{2} & = &
\int_{a}^{1-b} \int_{a}^{1-b}
\big( \min \{ v, w \} - v w \big) \;
\mbox{d} h_Y \big( G_Y^{-1}(v) \big) \,
\mbox{d} h_Y \big( G_Y^{-1}(w) \big)
\\[1ex]
 & = & \alpha^{-2}
\int_{a}^{1-(d/u)^{\alpha}} \int_{a}^{1-(d/u)^{\alpha}}
\big( \min \{ v, w \} - v w \big) \;
\mbox{d} \log (1-v) \, \mbox{d} \log (1-w)
\\[1ex]
 & \; =: & \alpha^{-2} J_t (a,1-(d/u)^{\alpha}; a, 1-(d/u)^{\alpha})
\end{eqnarray*}
and
\begin{eqnarray*}
(1-a-b)^2 \sigma_{11(3)}^{2} & = &
\int_{a}^{1-b} \int_{a}^{1-b}
\big( \min \{ v, w \} - v w \big) \;
\mbox{d} h_Y \big( G_Y^{-1}(v) \big) \,
\mbox{d} h_Y \big( G_Y^{-1}(w) \big)
\\[1ex]
 & = & \alpha^{-2}
\int_{a}^{1-b} \int_{a}^{1-b}
\big( \min \{ v, w \} - v w \big) \;
\mbox{d} \log (1-v) \, \mbox{d} \log (1-w)
\\[1ex]
 & = & \alpha^{-2} J_t (a,1-b; a, 1-b).
\end{eqnarray*}

Now, as follows from Theorem 3.1, the asymptotic variances of these two 
estimators of $\alpha$ are equal to $n^{-1} d_{11}^{(k)} \sigma^2_{11(k)} 
d_{11}^{(k)}$ for $k = 2,3$. This implies that the estimators 
$\widehat{\alpha}_{\mbox{\tiny $T$}}^{(2)}$, found by numerically 
solving (\ref{PaIyT2}), and $\widehat{\alpha}_{\mbox{\tiny $T$}}^{(3)}$, 
given by (\ref{PaIyT3}), have the following asymptotic distributions:
\begin{equation}
\widehat{\alpha}_{\mbox{\tiny $T$}}^{(2)} ~~is~~ 
{\cal{AN}}
\left( 
\alpha, \, \frac{\alpha^2}{n} \, 
\frac{J_t (a,1-(d/u)^{\alpha}; a,1-(d/u)^{\alpha})}{I_t^2(a,1-(d/u)^{\alpha})}
\right)
\label{PaIyT2-an}
\end{equation}
and
\begin{equation}
\widehat{\alpha}_{\mbox{\tiny $T$}}^{(3)} ~~is~~ 
{\cal{AN}}
\left( 
\alpha, \, \frac{\alpha^2}{n} \, 
\frac{J_t(a,1-b; a,1-b)}{I_t^2(a,1-b)}
\right).
\label{PaIyT3-an}
\end{equation}
From (\ref{PaIyT2-an}) we see that the asymptotic variance of 
$\widehat{\alpha}_{\mbox{\tiny $T$}}^{(2)}$ does not depend on the upper 
trimming proportion $b$, where $\frac{F(u)-F(d)}{1-F(d)} \leq 1-b \leq 1$. 
As expected, both estimators and their asymptotic distributions coincide 
when $1-b = \frac{F(u)-F(d)}{1-F(d)} = 1-(d/u)^{\alpha}$. Thus, for all 
practical purposes $\widehat{\alpha}_{\mbox{\tiny $T$}}^{(3)}$ is a better 
estimator (i.e., it has explicit formula and it becomes equivalent to
$\widehat{\alpha}_{\mbox{\tiny $T$}}^{(2)}$ if one chooses 
$b = (d/u)^{\alpha}$); therefore $\widehat{\alpha}_{\mbox{\tiny $T$}}^{(2)}$ 
(more generally, Case 2) will be discarded from further consideration.

As discussed in Note 3.3, the $T$-estimators are globally robust if $a > 0$ 
and $b > 0$. This is achieved by sacrificing the estimator's efficiency 
(i.e., the more robust the estimator the larger is its variance). From 
(\ref{PaImle-an}) and (\ref{PaIyT3-an}), we find that the asymptotic 
relative efficiency of $\widehat{\alpha}_{\mbox{\tiny $T$}}^{(3)}$ with 
respect to $\widehat{\alpha}_{\mbox{\tiny MLE}}$ is:
\[
\mbox{ARE} 
\left( 
\widehat{\alpha}_{\mbox{\tiny $T$}}^{(3)}, 
\widehat{\alpha}_{\mbox{\tiny MLE}}
\right) =~ 
\frac{\frac{\alpha^2}{n} \, \frac{1}{1 - ( d/u )^{\alpha}}}
{\frac{\alpha^2}{n} \, \frac{J_t(a,1-b; a,1-b)}{I_t^2(a,1-b)}} 
~=~
\frac{I_t^2(a,1-b)}{[ 1 - ( d/u )^{\alpha} ] J_t(a,1-b; a,1-b)} \, .
\]
In this case the integrals $I_t$ and $J_t$ can be derived analytically, but 
in general it is easier and faster to approximate them numerically; see 
Appendix A.2 in \cite{MR2591318} for specific approximation 
formulas of the bivariate integrals $J_t$. In Table 4.1, we present ARE 
computations.

\begin{center}
{\sc Table 4.1.} ~$\mbox{ARE} 
\left( \widehat{\alpha}_{\mbox{\tiny $T$}}^{(3)}, 
\widehat{\alpha}_{\mbox{\tiny MLE}} \right)$ 
for selected $a$ and $b$ and various choices of
\\[-1ex]
right-censoring proportion 
$\delta = 1 - \frac{F(u)-F(d)}{1-F(d)} = (d/u)^{\alpha}$.

\medskip

\begin{tabular}{c|ccccc|cccc|ccc}
\hline
\multicolumn{1}{c|}{} &
\multicolumn{5}{|c|}{$b$ (when $\delta = 0.01$)} &
\multicolumn{4}{|c|}{$b$ (when $\delta = 0.05$)} &
\multicolumn{3}{|c}{$b$ (when $\delta = 0.10$)} \\[-0.5ex]
\cline{2-13}
$a$ & 0.01 & 0.05 & 0.10 & 0.15 & 0.25 & 
0.05 & 0.10 & 0.15 & 0.25 & 0.10 & 0.15 & 0.25 \\
\hline
\hline
0 & 0.992 & 0.927 & 0.856 & 0.791 & 0.673 & 
0.966 & 0.892 & 0.824 & 0.701 & 0.941 & 0.870 & 0.740 \\
0.05 & 0.992 & 0.927 & 0.856 & 0.791 & 0.674 & 
0.966 & 0.892 & 0.825 & 0.702 & 0.942 & 0.871 & 0.741 \\
0.10 & 0.991 & 0.927 & 0.857 & 0.793 & 0.678 & 
0.966 & 0.893 & 0.826 & 0.704 & 0.943 & 0.872 & 0.744 \\
0.15 & 0.991 & 0.928 & 0.858 & 0.795 & 0.679 & 
0.967 & 0.894 & 0.828 & 0.708 & 0.944 & 0.874 & 0.747 \\
0.25 & 0.988 & 0.927 & 0.860 & 0.798 & 0.686 & 
0.966 & 0.896 & 0.832 & 0.715 & 0.946 & 0.878 & 0.755 \\
\hline
\end{tabular}
\end{center}

\bigskip

\noindent
It is obvious from the table that for a fixed $b$, the effect of 
the lower trimming proportion $a$ on the ARE is negligible. As $b$
increases, $T$-estimators become more robust but less efficient, 
yet their AREs are still sufficiently high (all at least 0.67;
more than half above 0.85). 
Also, all estimators' efficiency improves as the proportion of 
right-censored data $\delta$ increases. Take, for example, 
$a=b=0.10$: the $T$-estimator's efficiency grows from 0.857 
(when $\delta = 0.01$) to 0.943 (when $\delta = 0.10$).

\subsubsection{Payments {\em Z}}

Let $z_{1:n} \leq \cdots \leq z_{n:n}$ denote an ordered realization of 
variables (\ref{p2data}) with qf (\ref{p2qf}), where $F$ and $F^{-1}$ 
are given by (\ref{PaIcdf}) and (\ref{PaIqf}), respectively. Payments 
$Z$ are left- and right-censored and it follows from Note 3.2 that 
there are six permutations possible between the trimming proportions 
$a$, $b$ and $F(d)$, $F(u)$. However, analysis similar to the one done
in Section 4.2.1 shows that two of those scenarios (estimation based 
on censored data only) have no $\alpha$ in the formulas of population 
$T$ moments and three (estimation based on observed and censored data)
are inferior to the estimation scenario based on fully observed data. 
(Due to space limitations those investigations will not be presented 
here.) Thus, from now on we will focus on the following arrangement:
\[
0 \leq F(d) \leq a < 1-b \leq F(u) \leq 1.
\]

Similar to the previous section, standard empirical estimates of 
$F(d)$ and $F(u)$ provide guidance about the choice of $a$ and 
$1-b$. However, the function $h$ is defined differently: 
$h_Z(z) = \log(z/c+d)$. For Pareto I only the first $T$ moment is 
needed, and it is equal:
\begin{eqnarray*}
(1-a-b) \, T_{1(z)}(\alpha) & = & \int_a^{1-b} 
h_Z \left( G_Z^{-1} (v \, | \, \alpha) \right) \, dv
~=~ \int_a^{1-b} \log (F^{-1}(v)) \, dv
\\[1ex]
 & = & (1-a-b) \log (x_0) + \alpha^{-1} I_t(a, 1-b). 
\end{eqnarray*}
Matching $T_{1(z)}(\alpha)$ expression with $\widehat{T}_{1(z)} = 
(n-m_n-m_n^*)^{-1} \sum_{i = m_n + 1}^{n - m_n^*} 
\log( z_{i:n}/c + d )$ yields an explicit formula for a $T$-estimator
of $\alpha$:
\begin{equation}
\widehat{\widehat \alpha}_{\mbox{\tiny $T$}} = 
\frac{I_t(a,1-b)}{(1-a-b) [ \widehat{T}_{1(z)} - \log (x_0) ]} 
~=:~ s (\widehat{T}_{1(z)}).
\label{PaIzT}
\end{equation}

To specify the asymptotic distribution and compute AREs of 
$\widehat{\widehat \alpha}_{\mbox{\tiny $T$}}$, we again rely on 
Theorem 3.1. The single Jacobian entry for estimator (\ref{PaIzT})
is given by
\[
d_{11} = \frac{\partial \widehat{\widehat \alpha}_{\mbox{\tiny $T$}}}
{\partial \widehat{T}_{1(z)}} \Bigg|_{\widehat{T}_{1(z)} = T_{1(z)}} ~=~
\frac{\partial s (\widehat {T}_{1(z)})}
{\partial \widehat{T}_{1(z)}} \Bigg|_{\widehat{T}_{1(z)} = T_{1(z)}} ~=~ 
- \frac{(1-a-b) \alpha^2}{I_t(a,1-b)}.
\]
The single covariance-variance matrix entry, $\sigma_{11}^2$, 
is found as before:
\[
(1-a-b)^2 \sigma_{11}^{2} ~=~ \alpha^{-2} J_t (a,1-b; a, 1-b).
\]
Hence, the estimator $\widehat{\widehat \alpha}_{\mbox{\tiny $T$}}$,
given by (\ref{PaIzT}), has the following asymptotic distribution:
\begin{equation}
\widehat{\widehat \alpha}_{\mbox{\tiny $T$}} ~~is~~ 
{\cal{AN}}
\left( 
\alpha, \, \frac{\alpha^2}{n} \, 
\frac{J_t(a,1-b; a,1-b)}{I_t^2(a,1-b)}
\right).
\label{PaIzT-an}
\end{equation}

Now, from (\ref{PaImle2-an}) and (\ref{PaIzT-an}) we find that ARE 
of $\widehat{\widehat \alpha}_{\mbox{\tiny $T$}}$ with respect to 
$\widehat{\widehat \alpha}_{\mbox{\tiny MLE}}$ is
\begin{eqnarray*}
\mbox{ARE} 
\left( 
\widehat{\widehat \alpha}_{\mbox{\tiny $T$}}, 
\widehat{\widehat \alpha}_{\mbox{\tiny MLE}}
\right) 
& = &  
\frac{\frac{\alpha^2}{n} \, 
\left[ \frac{(x_0/d)^{\alpha}}{1-(x_0/d)^{\alpha}}
\log^2 \left[ (x_0/d)^{\alpha} \right] + 
(x_0/d)^{\alpha} - (x_0/u)^{\alpha} \right]^{-1}}
{\frac{\alpha^2}{n} \, \frac{J_t(a,1-b; a,1-b)}{I_t^2(a,1-b)}} 
\\[1ex]
 & = & 
\frac{I_t^2(a,1-b)}
{ \left[ \frac{(x_0/d)^{\alpha}}{1-(x_0/d)^{\alpha}}
\log^2 \left[ (x_0/d)^{\alpha} \right] + 
(x_0/d)^{\alpha} - (x_0/u)^{\alpha} \right] J_t(a,1-b; a,1-b)} \, .
\end{eqnarray*}
In Table 4.2, we present ARE computations for selected scenarios of 
data censoring.

\begin{center}
{\sc Table 4.2.} ~$\mbox{ARE} 
\left( \widehat{\widehat \alpha}_{\mbox{\tiny $T$}}, 
\widehat{\widehat \alpha}_{\mbox{\tiny MLE}} \right)$ 
for selected $a$ and $b$ and various combinations of left- and
\\[-1ex]
right-censoring proportions ($\delta_l, \delta_r$),
where 
$\delta_l = F(d) = 1-(x_0/d)^{\alpha}$ and
$\delta_r = 1-F(u) = (x_0/u)^{\alpha}$.

\medskip

{\small
\begin{tabular}{cc|ccccc|cccc|ccc}
\hline
\multicolumn{2}{c|}{} &
\multicolumn{5}{|c|}{$b$ (when $\delta_r = 0.01$)} &
\multicolumn{4}{|c|}{$b$ (when $\delta_r = 0.05$)} &
\multicolumn{3}{|c}{$b$ (when $\delta_r = 0.10$)} \\[-0.5ex]
\cline{3-14}
$\delta_l$ & $a$ & 0.01 & 0.05 & 0.10 & 0.15 & 0.25 & 
0.05 & 0.10 & 0.15 & 0.25 & 0.10 & 0.15 & 0.25 \\
\hline
\hline
0.50 & 0.50 & 0.973 & 0.923 & 0.864 & 0.809 & 0.708 & 
0.962 & 0.901 & 0.843 & 0.739 & 0.952 & 0.891 & 0.781 \\
 & 0.60 & 0.939 & 0.896 & 0.843 & 0.793 & 0.700 & 
0.934 & 0.879 & 0.827 & 0.730 & 0.929 & 0.874 & 0.772 \\
 & 0.70 & 0.882 & 0.849 & 0.805 & 0.761 & 0.679 & 
0.886 & 0.839 & 0.794 & 0.708 & 0.887 & 0.839 & 0.748 \\
 & 0.80 & 0.787 & 0.770 & 0.737 & 0.702 & -- & 
0.803 & 0.768 & 0.732 & -- & 0.812 & 0.774 & -- \\
\hline
0.75 & 0.75 & 0.927 & 0.898 & 0.855 & 0.811 & -- & 
0.941 & 0.895 & 0.850 & -- & 0.952 & 0.903 & -- \\
 & 0.80 & 0.868 & 0.848 & 0.812 & 0.773 & -- & 
0.889 & 0.850 & 0.810 & -- & 0.904 & 0.861 & -- \\
 & 0.85 & 0.789 & 0.781 & 0.753 & -- & -- & 
0.818 & 0.789 & -- & -- & 0.839 & -- & -- \\
\hline
0.85 & 0.85 & 0.896 & 0.887 & 0.856 & -- & -- & 
0.936 & 0.902 & -- & -- & 0.968 & -- & -- \\
 & 0.89 & 0.800 & 0.804 & 0.782 & -- & -- & 
0.848 & 0.825 & -- & -- & 0.886 & -- & -- \\
\hline
\end{tabular}
}
\end{center}

\bigskip

\noindent
Patterns in Table 4.2 are similar to those in Table 4.1, but in 
this case we also observe that $T$-estimators become more efficient 
as one or both censoring proportions increase. Take, for example, 
$a=0.80$ and $b=0.10$: the $T$-estimator's efficiency grows from
0.737 ($\delta_l = 0.50$, $\delta_r = 0.01$) to
0.812 ($\delta_l = 0.50$, $\delta_r = 0.10$) 
or from
0.768 ($\delta_l = 0.50$, $\delta_r = 0.05$) to
0.850 ($\delta_l = 0.75$, $\delta_r = 0.05$).

\subsection{{\em W}-Estimators}

As is evident from \eqref{Ts} and \eqref{Ws}, the ``central'' part 
of winsorized data is equal to trimmed data times $1-a-b$. Therefore, 
$W$-estimators will be closely related to the corresponding $T$-estimators. 
Choosing the same $h$ functions and trimming/winsorizing scenarios as in 
Section 4.2, we are able to derive $W$-estimators of $\alpha$ and their 
asymptotic distributions in a straightforward fashion.

\subsubsection{Payments {\em Y}}

Let $y_{1:n} \leq \cdots \leq y_{n:n}$ denote an ordered realization 
of $Y$ payments, $h_Y(y) = \log(y/(cd)+1)$, and 
$0 \leq a < 1-b \leq \frac{F(u)-F(d)}{1-F(d)} \leq 1$.
The population $W$-moment $W_{1(y)}(\alpha)$, given by equation 
(\ref{Wp}), is related to $T_{1(y)}(\alpha)$ and equal to
\begin{eqnarray*}
W_{1(y)}(\alpha) & = & 
   a \left[ h_Y \left( G_{Y}^{-1}(a \, | \, \alpha) \right) \right]
 + \int_a^{1-b} h_Y \left( G_Y^{-1} (v \, | \, \alpha) \right) \, dv
 + b \left[ h_Y \left( G_{Y}^{-1}(1-b \, | \, \alpha) \right) \right] 
\\[1ex]
 & = & a \left[ -\alpha^{-1} \log{(1-a)} \right] 
 + \alpha^{-1} I_t(a,1-b) + b \left[ -\alpha^{-1} \log{b} \right] 
\\[1ex]
 & = & \alpha^{-1} \left[ 1-a-b -\log (1-a) \right]
 ~=:~ \alpha^{-1} I_w(a,1-b).
\end{eqnarray*}
Matching $W_{1(y)}(\alpha)$ with the empirical $W$-moment
\[
\widehat{W}_{1(y)} = n^{-1}
\Big[
m_{n} \log{\big( y_{m_n+1:n}/(cd)+1 \big)} + 
\sum_{i=m_{n}+1}^{n-m_{n}^{*}}\log{\big( y_{i:n}/(cd)+1 \big)} + 
m_{n}^{*}\log{\big( y_{n-m_{n}^{*}:n}/(cd)+1 \big)}
\Big]
\]
yields an explicit formula for a $W$-estimator of $\alpha$:
\begin{equation}
\widehat{\alpha}_{\mbox{\tiny $W$}} ~=~
\frac{I_w(a,1-b)}{\widehat{W}_{1(y)}} ~=:~ r_y (\widehat{W}_{1(y)}).
\label{PaIyW}
\end{equation}
The asymptotic distribution of $\widehat{\alpha}_{\mbox{\tiny $W$}}$
follows from Theorem 3.2. The single Jacobian entry for estimator 
(\ref{PaIyW}) is given by
\[
d_{11} =
\frac{\partial \widehat{\alpha}_{\mbox{\tiny $W$}}}{\partial \widehat{W}_{1(y)}}
\Bigg|_{\widehat{W}_{1(y)}=W_{1(y)}}
=~ 
\frac{\partial r_y (\widehat{W}_{1(y)})}{\partial \widehat{W}_{1(y)}}
\Bigg|_{\widehat{W}_{1(y)}=W_{1(y)}}
=~
- \frac{\alpha^2}{I_w(a,1-b)}.
\]
The entry $\sigma_{11}^2$ is equal to 
$\widehat{A}_{1,1}^{(1)} + \cdots + \widehat{A}_{1,1}^{(4)}$
\citep[see Lemma A.1 in][]{MR3758788},
where 
$\widehat{A}_{1,1}^{(1)}, \ldots, \widehat{A}_{1,1}^{(4)}$ are 
derived as follows. Given that $\Delta_1 = 
W_{1(y)}(\alpha) = \alpha^{-1} \big( 1-a-b - \log (1-a) \big)$,
\begin{eqnarray*}
H_{1}(v) & = & h_Y \left( G_{Y}^{-1}(v) \right) ~=~ 
\log \left( \frac{G_{Y}^{-1}(v \, | \, \alpha)}{cd}+1 \right) 
\\[1ex]
 & = & - \alpha^{-1} \log (1-v) \, 
\mbox{\large\bf 1} \left\{ 0 \leq v < \frac{F(u)-F(d)}{1-F(d)} \right\}
 + \log (u/d) \,
\mbox{\large\bf 1} \left\{ \frac{F(u)-F(d)}{1-F(d)} \leq v \leq 1 \right\},
\end{eqnarray*}
and $H_{1}^{'}(v) = \alpha^{-1} (1-v)^{-1}
\mbox{\large\bf 1} \left\{ 0 < v < \frac{F(u)-F(d)}{1-F(d)} \right\}$,
we have:
\begin{eqnarray*}
\widehat{A}_{1,1}^{(1)} & = & \alpha^{-2} J_t (a,1-b; a,1-b),
\\[0.5ex]
\widehat{A}_{1,1}^{(2)} ~=~ \widehat{A}_{1,1}^{(3)} & = & 
\alpha^{-2} \left[ (1-a-b) \left( \frac{a^2}{1-a} - b \right) + 
b \log(1-a) - b \log b \right],
\\[0.5ex]
\widehat{A}_{1,1}^{(4)} & = & 
\alpha^{-2} \left[ \frac{a^2}{1-a}(a+2b) + b(1-b) \right].
\end{eqnarray*}
This yields
\begin{eqnarray*}
\sigma_{11}^{2} & = & \alpha^{-2}
\left[
J_t (a,1-b; a,1-b) + \frac{a^2(2-a)}{1-a} - 
b \big[ 1-2a-b + 2 \log b - 2 \log(1-a) \big]
\right] 
\\[0.5ex]
 & \, =: & \alpha^{-2} J_w (a,1-b; a,1-b).
\end{eqnarray*}
Putting it all together, $\widehat{\alpha}_{\mbox{\tiny $W$}}$, given 
by (\ref{PaIyW}), has the following asymptotic distribution:
\begin{equation}
\widehat{\alpha}_{\mbox{\tiny $W$}} ~~is~~ 
{\cal{AN}}
\left( 
\alpha, \, \frac{\alpha^2}{n} \, 
\frac{J_w(a,1-b; a,1-b)}{I_w^2(a,1-b)}
\right).
\label{PaIyW-an}
\end{equation}
Consequently,
\[
\mbox{ARE} 
\left( 
\widehat{\alpha}_{\mbox{\tiny $W$}}, 
\widehat{\alpha}_{\mbox{\tiny MLE}}
\right) =~ 
\frac{\frac{\alpha^2}{n} \, \frac{1}{1 - ( d/u )^{\alpha}}}
{\frac{\alpha^2}{n} \, \frac{J_w(a,1-b; a,1-b)}{I_w^2(a,1-b)}} 
~=~
\frac{I_w^2(a,1-b)}{[ 1 - ( d/u )^{\alpha} ] J_w(a,1-b; a,1-b)} \, .
\]
In Table 4.3, we present ARE computations for selected scenarios of 
data censoring.

\begin{center}
{\sc Table 4.3.} ~$\mbox{ARE} 
\left( \widehat{\alpha}_{\mbox{\tiny $W$}}, 
\widehat{\alpha}_{\mbox{\tiny MLE}} \right)$ 
for selected $a$ and $b$ and various choices of
\\[-1ex]
right-censoring proportion 
$\delta = 1 - \frac{F(u)-F(d)}{1-F(d)} = (d/u)^{\alpha}$.

\medskip

\begin{tabular}{c|ccccc|cccc|ccc}
\hline
\multicolumn{1}{c|}{} &
\multicolumn{5}{|c|}{$b$ (when $\delta = 0.01$)} &
\multicolumn{4}{|c|}{$b$ (when $\delta = 0.05$)} &
\multicolumn{3}{|c}{$b$ (when $\delta = 0.10$)} \\[-0.5ex]
\cline{2-13}
$a$ & 0.01 & 0.05 & 0.10 & 0.15 & 0.25 & 
0.05 & 0.10 & 0.15 & 0.25 & 0.10 & 0.15 & 0.25 \\
\hline
\hline
0 & 1.000 & 0.960 & 0.909 & 0.859 & 0.758 & 
1.000 & 0.947 & 0.895 & 0.789 & 1.000 & 0.944 & 0.833 \\
0.05 & 1.000 & 0.960 & 0.909 & 0.859 & 0.758 & 
1.000 & 0.947 & 0.895 & 0.789 & 1.000 & 0.944 & 0.833 \\
0.10 & 1.000 & 0.959 & 0.909 & 0.858 & 0.757 &
1.000 & 0.947 & 0.894 & 0.789 & 1.000 & 0.944 & 0.833 \\
0.15 & 0.999 & 0.958 & 0.908 & 0.857 & 0.756 &
0.999 & 0.946 & 0.893 & 0.788 & 0.999 & 0.943 & 0.832 \\
0.25 & 0.994 & 0.954 & 0.903 & 0.853 & 0.752 &
0.994 & 0.941 & 0.889 & 0.784 & 0.994 & 0.938 & 0.827 \\
\hline
\end{tabular}
\end{center}

\bigskip

\noindent
Patterns in Tables 4.1 and 4.3 are identical. However, it is worthwhile 
noting that for a fixed censoring scenario and fixed $a$ and $b$, each 
$W$-estimator is slightly more efficient than its $T$ counterpart.

\subsubsection{Payments {\em Z}}

Let $z_{1:n} \leq \cdots \leq z_{n:n}$ denote an ordered realization 
of $Z$ payments, $h_Z(z) = \log(z/c + d)$, and 
$0 \leq F(d) \leq a < 1-b \leq F(u) \leq 1$.
Then the population $W$-moment is equal to
\begin{eqnarray*}
W_{1(z)}(\alpha) & = & 
   a \left[ h_Z \left( G_{Z}^{-1}(a \, | \, \alpha) \right) \right]
 + \int_a^{1-b} h_Z \left( G_Z^{-1} (v \, | \, \alpha) \right) \, dv
 + b \left[ h_Z \left( G_{Z}^{-1}(1-b \, | \, \alpha) \right) \right] 
\\[1ex]
 & = & 
   a \left[ \log x_0 - \alpha^{-1} \log(1-a) \right]
 + (1-a-b) \log x_0 + \alpha^{-1} I_t(a,1-b)
 + b \left[ \log x_0 - \alpha^{-1} \log b \right] 
\\[1ex]
 & = & \log x_0 + \alpha^{-1} I_w(a,1-b).
\end{eqnarray*}
Matching $W_{1(z)}(\alpha)$ with the empirical $W$-moment
\[
\widehat{W}_{1(z)} = n^{-1}
\Big[
m_{n}\log{(z_{m_{n}+1:n}/c+d)}
+ \sum_{i=m_{n}+1}^{n-m_{n}^{*}}\log{(z_{i:n}/c+d)} 
+ m_{n}^{*}\log{(z_{n-m_{n}^{*}:n}/c+d)}
\Big]
\]
yields an explicit formula for a $W$-estimator of $\alpha$:
\begin{equation}
\widehat{\widehat \alpha}_{\mbox{\tiny $W$}} ~=~
\frac{I_w(a,1-b)}{\widehat{W}_{1(z)} - \log x_0} ~=:~ r_z (\widehat{W}_{1(z)}).
\label{PaIzW}
\end{equation}
The asymptotic distribution of $\widehat{\widehat \alpha}_{\mbox{\tiny $W$}}$
is derived by following the same steps as in Section 4.3.1. That is:
\[
d_{11} =
\frac{\partial \widehat{\widehat \alpha}_{\mbox{\tiny $W$}}}{\partial \widehat{W}_{1(z)}}
\Bigg|_{\widehat{W}_{1(z)}=W_{1(z)}}
=~ 
\frac{\partial r_z (\widehat{W}_{1(z)})}{\partial \widehat{W}_{1(z)}}
\Bigg|_{\widehat{W}_{1(z)}=W_{1(z)}}
=~
- \frac{\alpha^2}{I_w(a,1-b)}.
\]
Then, given that $\Delta_1 = W_{1(z)}(\alpha) = \log x_0 + \alpha^{-1} I_w(a,1-b)$ 
and, for $0 \leq F(d) \leq a < 1-b \leq F(u) \leq 1$,
$H_{1}(v) = h_Z \left( G_{Z}^{-1}(v) \right) = \log x_0 - \alpha^{-1} \log (1-v)$,
$H_{1}^{'}(v) = \frac{1}{\alpha (1-v)}$, we have
\begin{eqnarray*}
\sigma_{11}^{2} & = & \alpha^{-2}
\left[
J_t (a,1-b; a,1-b) + \frac{a^2(2-a)}{1-a} - 
b \big[ 1-2a-b + 2 \log b - 2 \log(1-a) \big]
\right] 
\\[0.5ex]
 & = & \alpha^{-2} J_w (a,1-b; a,1-b).
\end{eqnarray*}
Hence, $\widehat{\widehat \alpha}_{\mbox{\tiny $W$}}$, given by (\ref{PaIzW}), 
has the following asymptotic distribution:
\begin{equation}
\widehat{\widehat \alpha}_{\mbox{\tiny $W$}} ~~is~~ 
{\cal{AN}}
\left( 
\alpha, \, \frac{\alpha^2}{n} \, 
\frac{J_w(a,1-b; a,1-b)}{I_w^2(a,1-b)}
\right).
\label{PaIzW-an}
\end{equation}
Consequently,
\begin{eqnarray*}
\mbox{ARE} 
\left( 
\widehat{\widehat \alpha}_{\mbox{\tiny $W$}}, 
\widehat{\widehat \alpha}_{\mbox{\tiny MLE}}
\right)
& = &  
\frac{\frac{\alpha^2}{n} \, 
\left[ \frac{(x_0/d)^{\alpha}}{1-(x_0/d)^{\alpha}}
\log^2 \left[ (x_0/d)^{\alpha} \right] + 
(x_0/d)^{\alpha} - (x_0/u)^{\alpha} \right]^{-1}}
{\frac{\alpha^2}{n} \, \frac{J_w(a,1-b; a,1-b)}{I_w^2(a,1-b)}} 
\\[1ex]
 & = & 
\frac{I_w^2(a,1-b)}
{ \left[ \frac{(x_0/d)^{\alpha}}{1-(x_0/d)^{\alpha}}
\log^2 \left[ (x_0/d)^{\alpha} \right] + 
(x_0/d)^{\alpha} - (x_0/u)^{\alpha} \right] J_w(a,1-b; a,1-b)} \, .
\end{eqnarray*}
In Table 4.4, we present ARE computations for selected scenarios of 
data censoring.

Patterns in Table 4.4 are similar to those in Table 4.2. However,
unlike the ARE results in Tables 4.1 and 4.3, for payment $Z$ 
comparison of the $W$-estimators versus the $T$-estimators shows 
that neither method outperforms the other all the time. Each type
of estimators can have a better ARE than the competitor but that 
depends on the choice of $a$ and $b$ (which also depends on 
$\delta_l$ and $\delta_r$).

\begin{center}
{\sc Table 4.4.} ~$\mbox{ARE} 
\left( \widehat{\widehat \alpha}_{\mbox{\tiny $W$}}, 
\widehat{\widehat \alpha}_{\mbox{\tiny MLE}} \right)$ 
for selected $a$ and $b$ and various combinations of left- and
\\[-1ex]
right-censoring proportions ($\delta_l, \delta_r$),
where 
$\delta_l = F(d) = 1-(x_0/d)^{\alpha}$ and
$\delta_r = 1-F(u) = (x_0/u)^{\alpha}$.

\medskip

{\small
\begin{tabular}{cc|ccccc|cccc|ccc}
\hline
\multicolumn{2}{c|}{} &
\multicolumn{5}{|c|}{$b$ (when $\delta_r = 0.01$)} &
\multicolumn{4}{|c|}{$b$ (when $\delta_r = 0.05$)} &
\multicolumn{3}{|c}{$b$ (when $\delta_r = 0.10$)} \\[-0.5ex]
\cline{3-14}
$\delta_l$ & $a$ & 0.01 & 0.05 & 0.10 & 0.15 & 0.25 & 
0.05 & 0.10 & 0.15 & 0.25 & 0.10 & 0.15 & 0.25 \\
\hline
\hline
 0.50 & 0.50 & 0.968 & 0.929 & 0.880 & 0.831 & 0.733 & 
0.969 & 0.917 & 0.866 & 0.765 & 0.969 & 0.915 & 0.808 \\
 & 0.60 & 0.930 & 0.893 & 0.847 & 0.801 & 0.710 &  
0.932 & 0.883 & 0.835 & 0.741 & 0.933 & 0.883 & 0.783 \\
 & 0.70 & 0.877 & 0.843 & 0.802 & 0.761 & 0.680 &
0.880 & 0.836 & 0.793 & 0.709 & 0.884 & 0.838 & 0.749 \\
 & 0.80 & 0.796 & 0.769 & 0.734 & 0.701 & -- & 
0.802 & 0.766 & 0.731 & -- & 0.809 & 0.772 & --\\
\hline
 0.75 & 0.75 & 0.927 & 0.893 & 0.851 & 0.809 & -- & 
0.935 & 0.891 & 0.848 & -- & 0.948 & 0.901 & -- \\
 & 0.80 & 0.878 & 0.847 & 0.809 & 0.772 & -- & 
0.887 & 0.848 & 0.809 & -- & 0.901 & 0.860 & -- \\
 & 0.85 & 0.812 & 0.785 & 0.753 & -- & -- &  
0.823 & 0.789 & -- &-- & 0.839 & -- & -- \\
\hline
 0.85 & 0.85 & 0.922 & 0.892 & 0.856 & -- & -- & 
0.941 & 0.902 & -- & -- & 0.968 & -- & -- \\
 & 0.89 & 0.838 & 0.814 & 0.783 & -- & -- & 
0.858 & 0.826 & -- & -- & 0.886 & -- & -- \\
\hline
\end{tabular}
}
\end{center}

\medskip

\section{Real Data Example}
\label{sec:RealDataExample}

In this section, we use MLE and several $T$ and $W$ estimators for 
fitting the Pareto I model to the well-studied Norwegian fire claims 
data 
(see 
\citealp{MR2035058};
\citealp{nb15};
\citealp{MR3474025};
\citealp{ann21}),
which are available at the following website:
\begin{center}
{\tt http://lstat.kuleuven.be/Wiley}
~ (in Chapter 1, file {\sc norwegianfire.txt}).
\end{center}

\subsection{Data and Preliminary Diagnostics} 

The data represent the total damage done by fires in Norway for the 
years 1972 through 1992; only damages in excess of a priority of 
500,000 Norwegian krones ({\sc nok}) are available. We will analyze 
the data set for the year 1975, which has $n=142$ observations with 
the most extreme loss of 52.6 million {\sc nok}. The data for this 
year was also modeled with Pareto I by \cite{MR2035058}. 
Table 5.1 provides a summary of the data set.

\newpage

\begin{center}
{\sc Table 5.1.} ~Summary of {\em Norwegian Fire Claims\/} data 
for the year 1975.

\medskip

\begin{tabular}{|c|cccccc|}
\hline
{\em Severity\/} (millions {\sc nok}) & 
$[0.5; \, 1.0)$ & $[1.0; \, 2.0)$ & $[2.0; \, 5.0)$ &
$[5.0; \, 10.0)$ & $[10.0; \, 20.0)$ &  $[20.0; \, \infty)$\\
\hline
{\em Relative Frequency} & 0.54 & 0.28 & 0.12 & 0.03 & 0.02 & 0.01 \\
\hline
\end{tabular}
\end{center}

\bigskip

Since no information is given below 500,000 and there is no policy limit
and coinsurance, the random variable that generated the data is related 
to payment $Y$, i.e., it is $Y+d$ with $c=1$, $d=500,000$, and $u=\infty$.
Moreover, as is evident from Table 5.1, the data are right-skewed and 
heavy-tailed suggesting that Pareto I, with cdf (\ref{PaIcdf}) and qf
(\ref{PaIqf}), might be an appropriate model in this case. To see how 
right-censoring changes the estimates of $\alpha$, model fits, and 
ultimately premium estimates for a layer, we will consider two data 
scenarios: {\em Original Data\/} ($c=1$, $d=500,000$, $u=\infty$) 
and {\em Modified Data\/} ($c=1$, $d=500,000$, $u=7,000,000$).

\begin{figure}[hbt!]
\centering
\includegraphics[width=0.47\textwidth]{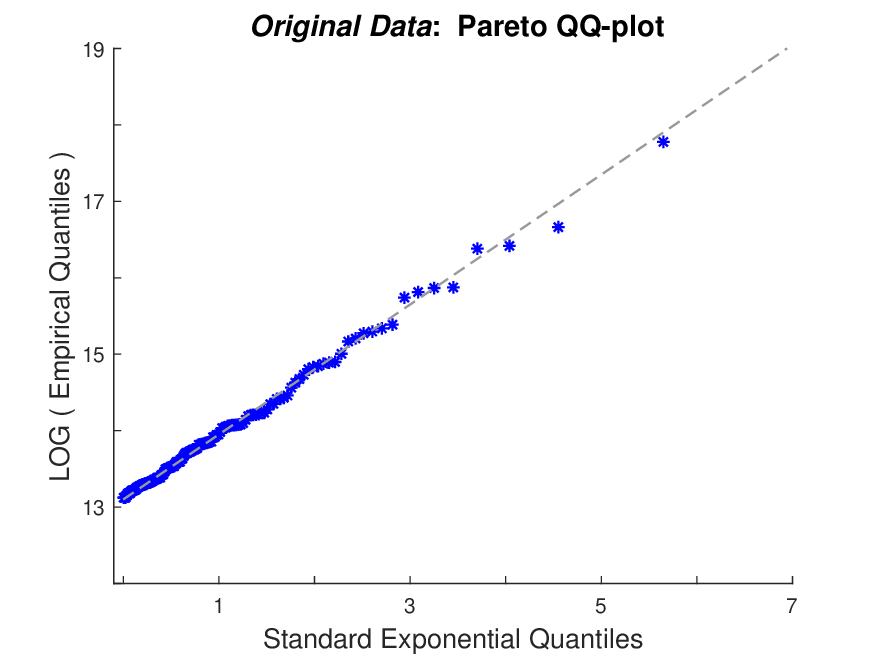}
\includegraphics[width=0.47\textwidth]{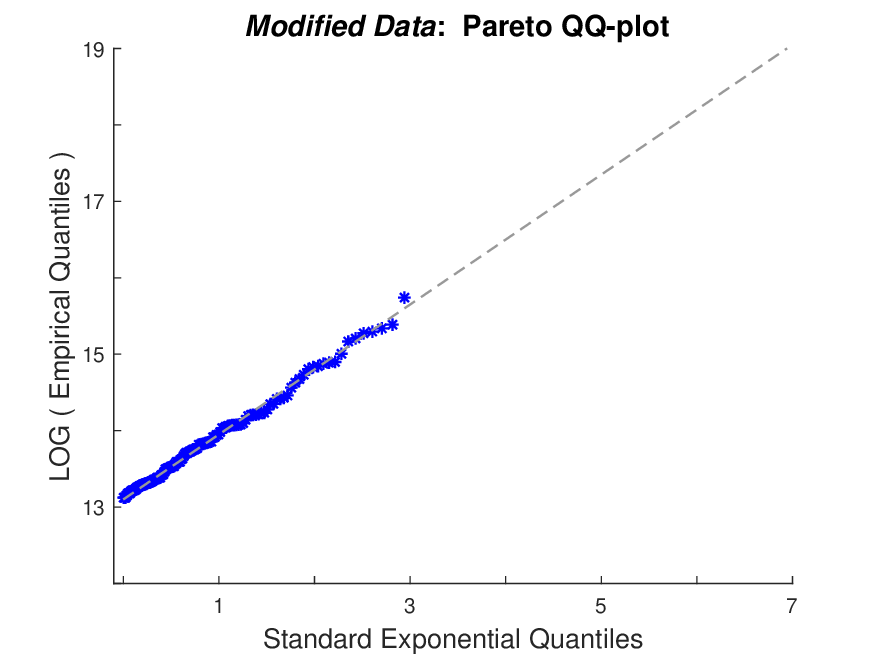}
\caption{Pareto quantile-quantile 
plots for the original  and modified data sets.
The dashed line represents the ``best'' fit line 
(in both cases): $y = 13.1 + 0.85 \, x$.}
\label{fig:QQ_Plot}
\end{figure}

Further, we will fit Pareto I$\; (x_0, \alpha)$ under the original 
and modified data scenarios. 
Preliminary diagnostics -- the quantile-quantile plots (QQ-plots) 
presented in Figure \ref{fig:QQ_Plot} -- strongly suggests that the Pareto I 
assumption is reasonable. Note that the plots are parameter free. 
That is, since Pareto I is a log-location-scale family, its QQ-plot 
can be constructed without first estimating model parameters. Note 
also that only actual data can be used in these plots (i.e., no 
observations $u = 7,000,000$ under the modified data scenario).

\subsection{Model Estimation and Validation} 

Compute parameter estimates $\widehat{\alpha}$ we use
the following formulas: (\ref{PaImle}) for MLE, (\ref{PaIyT3})
for $T$, and (\ref{PaIyW}) for $W$. In order to match the 
specifications of fire claims data (denoted $l_{1}, \ldots, l_{142}$), in (\ref{PaImle})
$c=1$ and $y_i+d$ is replaced with $l_i$; and in (\ref{PaIyT3}) and 
(\ref{PaIyW}) function $h_Y$ is now defined as $h_Y(l_i) = \log (l_i/d)$.
Specifically, for {\em modified data\/} ($d=0.5 \times 10^6$, 
$u=7 \times 10^6$, claims $l_1, \ldots, l_n$, $n=142$), 
MLE is given by
\[
\widehat{\alpha}_{\mbox{\tiny MLE}} = 
\frac{\sum_{i=1}^{n} \mbox{\large\bf 1} \{ d < l_i < u \}}
{\sum_{i=1}^{n} \log ( l_i/d ) \mbox{\large\bf 1} \{ d < l_i < u \} +
\log (u/d) \sum_{i=1}^{n} \mbox{\large\bf 1} \{ l_i = u \}},
\]
and for {\em original data\/} ($d=0.5 \times 10^6$, $u=\infty$, 
claims $l_1, \ldots, l_{n}$, $n=142$), it becomes
$
\widehat{\alpha}_{\mbox{\tiny MLE}} = 
\frac{n}{\sum_{i=1}^{n} \log ( l_i/d )}.
$
Computational formulas for the $T$ and $W$ estimators remain 
the same for both data scenarios:
\[
\widehat{\alpha}_{\mbox{\tiny $T$}} = 
\frac{(1-a) (1 - \log (1-a)) - b (1 - \log b)}{(1-a-b) \, 
\widehat{T}_{1(y)}}
\qquad
\mbox{and}
\qquad
\widehat{\alpha}_{\mbox{\tiny $W$}} = 
\frac{1-a-b- \log (1-a)}{\widehat{W}_{1(y)}},
\]
where 
$
\widehat{T}_{1(y)} = (n-m_{n}-m_{n}^*)^{-1}
\sum_{i = m_{n} + 1}^{n - m_{n}^*} \log( l_i / d)
$
~and
\[
\widehat{W}_{1(y)} = n^{-1}
\Big[
m_{n} \log{\big( l_{m_{n}+1} / d \big)} + 
\sum_{i=m_{n}+1}^{n-m_{n}^{*}}\log{\big( l_{i}/ d \big)} + 
m_{n}^{*}\log{\big( l_{n-m_{n}^{*}}/ d \big)}
\Big],
\]
with several choices of $m_{n} = [n a]$ and $m_{n}^{*} = [n b]$.
The corresponding asymptotic distributions are specified by 
(\ref{PaImle-an}), (\ref{PaIyT3-an}), and (\ref{PaIyW-an}). They
are used to construct the 90\% confidence intervals for $\alpha$.
All computations are summarized in Table 5.2, where goodness-of-fit 
analysis is also provided;
see \cite{MR3890025} for how to perform Kolmogorov-Smirnov 
(KS) test for right-censored data (Section 15.4.1) and how to estimate 
its $p$-value using parametric bootstrap (Section 19.4.5).

\begin{center}
{\sc Table 5.2.} ~Pareto I$\; (x_0 = 7000, \alpha)$ fitted 
to the original and modified data sets. Point and
\\[-2mm]
90\% confidence interval estimates of $\alpha$, Kolmogorov-Smirnov 
(KS) statistics and their $p$-values.

\medskip

\begin{tabular}{c|cc|cc|cc|cc}
\hline
\multicolumn{1}{c|}{Estimator} &
\multicolumn{4}{|c|}{\em Original Data} &
\multicolumn{4}{|c}{\em Modified Data} \\[-0.25ex]
\cline{2-9}
 & $\widehat{\alpha}$ & 90\% CI & KS & $p$-value$^*$ &
$\widehat{\alpha}$ & 90\% CI & KS & $p$-value$^*$ \\
\hline\hline
MLE & 1.22 & [1.05; 1.39] & 0.05 & 0.70 & 1.20 & [1.03; 1.37] & 0.05 & 0.71 \\
\hline
$T, \, a = b = 0$ & 1.22 & [1.05; 1.39] & 0.05 & 0.70 & 
-- & -- & -- & -- \\
$T, \, a = b = 0.10$ & 1.22 & [1.04; 1.41] & 0.05 & 0.61 & 
1.22 & [1.04; 1.41] & 0.05 & 0.69 \\
$T, \, a = 0.05, b = 0.15$ & 1.22 & [1.03; 1.41] & 0.05 & 0.60 & 
1.22 & [1.03; 1.41] & 0.05 & 0.68 \\
\hline
$W, \, a = b = 0$ & 1.22 & [1.05; 1.39] & 0.05 & 0.70 & 
-- & -- & -- & -- \\
$W, \, a = b = 0.10$ & 1.22 & [1.04; 1.40] & 0.05 & 0.68 & 
1.22 & [1.04; 1.40] & 0.05 & 0.74 \\
$W, \, a = 0.05, b = 0.15$ & 1.21 & [1.03; 1.39] & 0.05 & 0.59 & 
1.21 & [1.03; 1.39] & 0.05 & 0.68 \\
\hline
\multicolumn{9}{c}{} \\[-2.75ex]
\multicolumn{9}{l}{$^*$ {\footnotesize The $p$-values are computed using 
parametric bootstrap with 1000 simulation runs.}} \\
\end{tabular}
\end{center}

\bigskip

As is evident from Table 5.2, all estimators exhibit excellent goodness-of-fit 
performance, as one would expect after examining Figure \ref{fig:QQ_Plot}. 
Irrespective of 
the method of estimation the fitted Pareto I model has very heavy right tail, 
i.e., for $1 < \alpha < 2$ all its moments are infinite except the mean. 
The $T$ and $W$ estimators with $a=b=0$ match the estimates of MLE under the 
original data scenario. As discussed in Section 4.2, this choice of $a$ and 
$b$, however, would be inappropriate when data are censored at $u = 7,000,000$, 
which corresponds to about 4.9\% of censoring. Clearly, this level of censoring 
has no effect whatsoever on $T$ and $W$ estimators with $a=b=0.10$ and $a=0.05, 
b=0.15$, which demonstrates their robustness. The MLE, on the other hand, is 
affected by censoring. While the change in its estimated values of $\alpha$ 
and the corresponding confidence intervals seems minimal (less than 2\%), 
it gets magnified when applied to calculation of premiums, as will be 
shown next.

\subsection{Contract Pricing} 

Let us consider estimation of the loss severity component 
of the {\em pure premium\/} for an insurance benefit ($B$) that equals 
to the amount by which a fire loss damage ($L$) exceeds 7 million 
{\sc nok} with a maximum benefit of 28 million {\sc nok}. That is,
\begin{equation}
\label{benefit}
B = \begin{cases}
   0, & \mbox{if} ~~ L \leq d^*; \\[0.25ex]
 L-d^*, & \mbox {if} ~~ d^* < L \leq u^*; 
\\[0.25ex]
  u^* - d^*, & \mbox {if} ~~ L > u^*,
\end{cases}
\end{equation}
and, if $L$ follows the distribution function $F$, we seek
\[
\varPi [F] ~=~ \mbox{\bf E}[B] ~=~ \int_{d^*}^{u^*} 
(x - d^*) \, dF(x) + (u^* - d^*) [ 1-F(u^*) ] ~=~
\int_{d^*}^{u^*} [ 1 - F(x) ] \, d x,
\]
where $d^* = 7 \cdot 10^6$ and $u^* = 35 \cdot 10^6$. (In U.S. 
dollars, this roughly corresponds to the layer from 1 to 5 million.)
We will present premium estimates for two versions of $L$: 
{\em Observed Loss\/} (corresponds to $L \sim \mbox{Pareto I} \; 
(d=5 \cdot 10^5, \alpha)$) 
and {\em Ground-Up Loss\/} (corresponds to $L \sim \mbox{Pareto I} \; 
(x_0=7 \cdot 10^3, \alpha)$). The second version shows how different 
the premium is if all -- observed and unobserved -- data were available. 
It also facilitates evaluation of various loss variable characteristics; 
for example, if one switches from a priority of 500,000 to 250,000, 
the change in loss elimination ratio could be estimated, but such 
computations are impossible under the first version of $L$.

Now, straightforward derivations yield the following expression for 
$\varPi [F]$:
\begin{equation}
\label{prem}
\varPi [F] ~=~ C \times \frac{(u^*/C)^{1-\alpha} - 
(d^*/C)^{1-\alpha}}{1-\alpha}, \qquad  \alpha \ne 1,
\end{equation}
where $C = d$ (for observed loss) or $= x_0$ (for ground-up loss).
If $\alpha=1$, then $\varPi [F] = C \log(u^*/d^*)$. To get point 
estimates $\varPi [\widehat{F}]$, we plug in the estimates of $\alpha$ 
from Table 5.2 into (\ref{prem}). To construct interval estimators, 
we rely on the delta method 
\citep[see][Section 3.3]{MR595165}, which 
uses the asymptotic distributions (\ref{PaImle-an}), (\ref{PaIyT3-an}), 
and (\ref{PaIyW-an}) and transforms $\widehat{\alpha}$ according to 
(\ref{prem}). Thus, we have that $\varPi [\widehat{F}]$ is asymptotically 
normal with mean $\varPi [F]$ and variance
$
\mbox{\bf Var} (\widehat{\alpha}) \times 
\left( 
\frac{\partial}{\partial \alpha} \Big[ \varPi [F] \Big] 
\right)^2,
$
where 

{\small
\[
\frac{\partial}{\partial \alpha} \Big[ \varPi [F] \Big] 
~=~ \frac{C}{(1-\alpha)^2} 
\left\{
(1-\alpha) 
\left[ 
\left( \frac{d^*}{C} \right)^{1-\alpha} \log \left( \frac{d^*}{C} \right) -
\left( \frac{u^*}{C} \right)^{1-\alpha} \log \left( \frac{u^*}{C} \right)
\right]
+ 
\left( \frac{u^*}{C} \right)^{1-\alpha} -
\left( \frac{d^*}{C} \right)^{1-\alpha}
\right\}
\]
}

\noindent
and $\mbox{\bf Var} (\widehat{\alpha})$ is taken from (\ref{PaImle-an}), 
(\ref{PaIyT3-an}), or (\ref{PaIyW-an}). To assure that the left endpoint 
of the confidence intervals is positive, we will construct log-transformed 
intervals which have the following structure: 
$\big[ \varPi [\widehat{F}] \cdot K^{-1}; \,
\varPi [\widehat{F}] \cdot K \big]$ for $K > 0$.
Table 5.3 presents point and 90\% log-transformed interval estimates 
of premiums for observed and ground-up losses under the original and 
modified data scenarios.

\begin{center}
{\sc Table 5.3.} ~~Point and 90\% log-transformed interval estimates of 
$\varPi [F]$
\\[-2mm] 
for observed and ground-up loss $L$ under the original and modified data 
scenarios.

\medskip

{\small
\begin{tabular}{c|cc|cc|cc|cc}
\hline
\multicolumn{1}{c|}{} &
\multicolumn{4}{|c|}{{\em Observed Loss} $(\times 10^5)$} &
\multicolumn{4}{|c}{\em Ground-up Loss $(\times 10^3)$} \\[-0.25ex]
\cline{2-9}
\multicolumn{1}{c|}{Estimator} &
\multicolumn{2}{|c|}{\em Original} &
\multicolumn{2}{|c|}{\em Modified} &
\multicolumn{2}{|c|}{\em Original} &
\multicolumn{2}{|c}{\em Modified} \\[-0.25ex]
\cline{2-9}
\multicolumn{1}{c|}{} &
\multicolumn{2}{|c|}{} &
\multicolumn{2}{|c|}{} &
\multicolumn{2}{|c|}{} &
\multicolumn{2}{|c}{} \\[-3ex]
 & $\varPi [\widehat{F}]$ & 90\% CI & 
 $\varPi [\widehat{F}]$ & 90\% CI & 
 $\varPi [\widehat{F}]$ & 90\% CI & 
 $\varPi [\widehat{F}]$ & 90\% CI \\
\hline\hline
MLE & 
3.82 & [2.16; 6.77] & 4.01 & [2.25; 7.14] & 
2.11 & [0.58; 7.67] & 2.35 & [0.64; 8.65] \\
\hline
$T, \, a = b = 0$ & 
3.82 & [2.16; 6.77] & -- & -- &
2.11 & [0.58; 7.67] & -- & -- \\
$T, \, a = b = 0.10$ & 
3.77 & [2.02; 7.01] & 3.77 & [2.02; 7.01] &
2.04 & [0.50; 8.32] & 2.04 & [0.50; 8.32] \\
$T, \, a = 0.05, b = 0.15$ & 
3.75 & [1.96; 7.17] & 3.75 & [1.96; 7.17] &
2.03 & [0.47; 8.75] & 2.03 & [0.47; 8.75] \\
\hline
$W, \, a = b = 0$ & 
3.82 & [2.16; 6.77] & -- & -- &
2.11 & [0.58; 7.67] & -- & -- \\
$W, \, a = b = 0.10$ & 
3.77 & [2.06; 6.89] & 3.77 & [2.06; 6.89] &
2.05 & [0.52; 8.00] & 2.05 & [0.52; 8.00] \\
$W, \, a = 0.05, b = 0.15$ & 
3.92 & [2.12; 7.26] & 3.92 & [2.12; 7.26] &
2.24 & [0.56; 8.99] & 2.24 & [0.56; 8.99] \\
\hline
\end{tabular}
}
\end{center}

\bigskip

As can be seen from Table 5.3, premiums for the ground-up loss are 
two orders of magnitude smaller than those for the observed loss.
This was expected because the ground-up distribution automatically
estimates that the number of losses below 500,000 is large while 
the observed loss distribution assumes that that number is zero.
Further, as the data scenario changes from original to modified,
the robust estimates of premiums ($T$ and $W$ with $a=b=0.10$ 
and $a=0.05$, $b=0.15$) do not change, but those based on MLE 
increase by 5\% (for observed loss) and 11\% (for ground-up loss).
Finally, note that these MLE-based premium changes occur albeit
Pareto I fits the data exceptionally well (see Table 5.2). 
If the model fits were less impressive, the premium swings would 
be more pronounced.

\subsection{Additional Illustrations}
\label{sec:AdditionalIllustrations}

It was mentioned in Section 1 that robust model fits can be achieved 
by other methods of estimation; one just needs to apply them to trimmed 
or winsorized data. Since for the Pareto I distribution $T$ and $W$ 
estimators of $\alpha$ with $a=b=0$ coincide with MLE (see Table 5.2), 
it is reasonable to expect that left- and/or right-censored MLE 
should behave like a $W$ estimator with similarly chosen winsorizing 
proportions. (This kind of strategy is sometimes used in data 
analysis practice to robustify MLE.) In what follows, we 
investigate how this idea works on Norwegian fire claims. 

First of all, the asymptotic properties of MLE as stated in Section 4.1 
are valid when the right-censoring threshold $u$ is fixed, hence 
the probability to exceed it is random. 
Fixed thresholds method of moments and its some
variants have been investigated by \cite{MR4192140}.
The corresponding properties
for $T$ and $W$ estimators are established under the complete opposite
scenario: data proportions are fixed but thresholds are random.
To see what effect this difference has on actual estimates of $\alpha$, 
we compute MLEs by matching its censoring points with those used for 
the $T$ and $W$ estimators in Table 5.2. In particular, for $a = b = 0.10$, 
we have $m_n = m_n^* = [14.2] = 14$ which implies that for observations 
from $l_{15} = 0.551 \cdot 10^6$ to $l_{128} = 3.289 \cdot 10^6$ their
actual values are included in the computation of 
$\widehat{\alpha}_{\mbox{\tiny $W$}}$ and for the remaining ones the 
minimum and maximum of actual observations are used, 
i.e., $l_1 = \cdots = l_{14} = 0.551 \cdot 10^6$ and 
$l_{129} = \cdots = l_{142} = 3.289 \cdot 10^6$.
When computing censored MLE, this kind of effect on data can be 
achieved by choosing the left- and right-censoring levels 
$\widetilde{d}$ and $\widetilde{u}$ as follows: 
$\widetilde{d} = 0.551 \cdot 10^6$ and 
$\widetilde{u} = 3.289 \cdot 10^6$.
Likewise, for $a = 0.05$ and $b = 0.15$, we have $m_n = [7.1] = 7$
and $m_n^* = [21.3] = 21$ and arrive at 
$\widetilde{d} = 0.530 \cdot 10^6$ and 
$\widetilde{u} = 2.497 \cdot 10^6$.
Note that $\widetilde{d}$ and $\widetilde{u}$ are {\em not fixed\/}, 
which is required for derivations of asymptotic properties, rather they 
are {\em estimated\/} threshold levels. Rigorous theoretical treatment 
of MLEs with estimated threshold levels is beyond the scope of the current 
paper and thus is deferred to future research projects. For illustrative 
purposes, however, we can {\em assume\/} that the threshold levels 
$\widetilde{d}$ and $\widetilde{u}$ are fixed and apply the methodology 
of Section 4.1.

Due to the left-truncation of Norwegian fire claims at $d = 500,000$ and
additional left- and right-censoring at $\widetilde{d}$ ($\widetilde{d} > d$) 
and $\widetilde{u}$, respectively, we are fitting Pareto I$\; (d, \alpha)$ 
to Payment $Z$ data. Given these modifications, 
$\widehat{\widehat{\alpha}}_{\mbox{\tiny MLE}}$ 
(censored at $\widetilde{d}$ and $\widetilde{u}$) is found by maximizing 
(\ref{PaImle2}) of the following form:
\begin{eqnarray*}
{\cal{L}}_{P_Z} \big( \alpha \, \big| \, l_1, \ldots, l_n \big) & = &  
\log \big[ 1 - (d/\widetilde{d})^{\alpha} \big] \sum_{i=1}^n 
\mbox{\large\bf 1} \{ l_i = \widetilde{d} \}
~+~ \alpha \log (d/\widetilde{u}) \sum_{i=1}^n 
\mbox{\large\bf 1} \{ l_i =  \widetilde{u} \}
\\
& & + ~ \sum_{i=1}^n \big[ \log \left( \alpha/d \right) - 
(\alpha+1) \log \left( l_i/d \right) \big]
\mbox{\large\bf 1} \{ \widetilde{d} < l_i < \widetilde{u} \}. \qquad ~
\end{eqnarray*}
Similarly, the asymptotic distribution (\ref{PaImle2-an}) 
should be of the following form:
\[
\widehat{\widehat{\alpha}}_{\mbox{\tiny MLE}} ~~is~~ 
{\cal{AN}}
\left( 
\alpha, \, \frac{\alpha^2}{n} \, 
\left[ 
\frac{(d/\widetilde{d})^{\alpha}}{1-(d/\widetilde{d})^{\alpha}}
\log^2 \left[ (d/\widetilde{d})^{\alpha} \right] + 
(d/\widetilde{d})^{\alpha} - (d/\widetilde{u})^{\alpha} \right]^{-1}
\right).
\]
Numerical implementation of these formulas is provided in Table 5.4, 
where we compare censored MLEs with $W$ estimators based on such 
$a$ and $b$ that act on data the same way as MLEs. It is clear 
from the table that censored MLEs do achieve the same degree of 
robustness as the corresponding $W$ estimators. Moreover, the point 
and interval estimates produced by these two methods are very close 
but not identical. Finally, it should be emphasized once again that
the MLE-based intervals are constructed using the assumed asymptotic 
distribution which is not proven and may be incorrect.

\begin{center}
{\sc Table 5.4.} ~Comparison of $W$'s and censored MLEs of 
$\alpha$ of Pareto I$\; (d = 500,000, \alpha)$ 
\\[-2mm]
fitted to the original and modified data sets. Note that 
$\widetilde{d}$ and $\widetilde{u}$ are {\em assumed\/} 
to be fixed.

\medskip

\begin{tabular}{c|cc|cc|cc}
\hline
\multicolumn{1}{c|}{Estimator} &
\multicolumn{2}{|c|}{Censoring Thresholds} &
\multicolumn{2}{|c|}{\em Original Data} &
\multicolumn{2}{|c}{\em Modified Data} \\[-0.25ex]
\cline{4-7}
& $\widetilde{d} \; (\times 10^6)$ & $\widetilde{u} \; (\times 10^6)$ &
$\widehat{\alpha}$ & 90\% CI & $\widehat{\alpha}$ & 90\% CI \\
\hline\hline
MLE & 0.551 & 3.289 & 1.2155 & 
[1.0385; 1.3925] & 
1.2155 & [1.0385; 1.3925] \\
$W, \, a = b = 0.10$ & $--$ & $--$ & 
1.2218 & [1.0440; 1.3996] & 1.2218 & [1.0440; 1.3996] \\
\hline
MLE & 0.530 & 2.497 & 1.2046 & [1.0249; 1.3843] & 
1.2046 & [1.0249; 1.3843] \\
$W, \, a = 0.05, b = 0.15$ & $--$ & $--$ & 
1.2099 & [1.0288; 1.3910] & 1.2099 & [1.0288; 1.3910] \\
\hline
\end{tabular}
\end{center}

\medskip

\section{Concluding Remarks}

In this paper, we have developed the methods of {\em trimmed\/} 
(called $T$) and {\em winsorized\/} (called $W$) moments for 
robust estimation of claim severity models that are affected by 
deductibles, policy limits and coinsurance. The definitions and 
asymptotic properties of these estimators have been provided
for various data scenarios, including {\em complete\/}, 
{\em truncated\/}, and {\em censored\/} data, and two types of 
{\em insurance payments\/}. Further, specific definitions and 
explicit asymptotic distributions of the maximum likelihood 
(MLE), $T$, and $W$ estimators have been derived for insurance 
payments when the loss variable follows a single-parameter 
Pareto distribution. These analytic examples have clearly shown 
that $T$ and $W$ estimators sacrifice little efficiency with 
respect to MLE, but are robust and have explicit formulas 
(whereas finding MLE does require numerical optimization; 
see Section 4.1.2). These are highly desirable properties in 
practice. Finally, the practical performance of the estimators 
under consideration have been illustrated using the well-known 
Norwegian fire claims data.

The research presented in this paper invites follow-up studies 
in several directions. For example, the most obvious direction
is to study small-sample properties of these estimators (for 
Pareto $\alpha$) using simulations. Second, to derive specific 
formulas and investigate the estimators' efficiency properties 
for other loss models such as lognormal, gamma, log-logistic, 
folded-$t$, and GB2 distributions. Third, to consider robust 
estimation based on different influence functions such as Hampel's 
redescending or Tukey's biweight (bisquare) functions. Fourth, 
to compare practical performance of our models' robustness with 
that based on model distance and entropy. 
Note that the latter 
approach derives the worst-case risk measurements, relative to 
measurements from a baseline model, and has been used by authors 
in the actuarial literature 
\citep[e.g.,][]{MR3928500}
as well as in the financial risk management literature 
(see, e.g., 
\citealp{as12};
\citealp{MR3175968}).
Fifth, it is also of interest to see how well future insurance
claims can be predicted using the robust parametric approach 
of this paper versus more general predictive techniques that 
are designed to incorporate model uncertainty 
(see, e.g.,
\citealp{MR3656190};
\citealp{hkm18}).

\baselineskip 4.80mm
\setlength{\bibsep}{5.0pt plus 0.00ex}
\bibliography{ArXiv}
\thispagestyle{plain}
\pagestyle{plain} 
\thispagestyle{plain}

\end{document}